\documentclass[10pt, doublecolumn]{IEEEtran}
\usepackage{epsfig,latexsym}
\usepackage{float}
\usepackage{indentfirst}
\usepackage{amsmath}
\usepackage{bm}
\usepackage{amssymb}
\usepackage{times}
\usepackage{enumitem}

\usepackage{algorithm}
\usepackage[noend]{algpseudocode}
\usepackage{subfigure}
\usepackage{psfrag}
\usepackage{hyperref}
\usepackage{cite}
\usepackage{lastpage}
\usepackage{fancyhdr}
\usepackage{color} 
 \usepackage{amsthm}
\usepackage{bigints}
\usepackage{array}
\usepackage{booktabs}
\newtheorem{Lemma}{Lemma}
\newtheorem{Corollary}{Corollary}
\newtheorem{lemma}[Lemma]{$\mathbf{Lemma}$}

\newtheorem{corollary}[Corollary]{$\mathbf{Corollary}$}

\newcounter{problem}
\newcounter{save@equation}
\newcounter{save@problem}
\makeatletter

\begin{document}
\title{  \vspace{-0.5em}\huge{  Flexible-Antenna Systems: A Pinching-Antenna Perspective    }}

\author{ Zhiguo Ding, \IEEEmembership{Fellow, IEEE}, Robert Schober, \IEEEmembership{Fellow, IEEE}, and H. Vincent Poor, \IEEEmembership{Life Fellow, IEEE}   \thanks{ 
  
\vspace{-2em}

Z. Ding is with Khalifa University, Abu Dhabi, UAE, and the University
of Manchester, Manchester, M1 9BB, UK.    
R. Schober is with the Institute for Digital Communications,
Friedrich-Alexander-University Erlangen-Nurnberg (FAU), Germany. H. V. Poor is  with the  Department of Electrical and Computer Engineering, Princeton University,
Princeton, NJ 08544, USA.
 

  }\vspace{-2.5em}}
 \maketitle

\begin{abstract}
Flexible-antenna systems have recently received significant research interest due to their capability to reconfigure wireless channels intelligently. This paper focuses on a new type of flexible-antenna technology, termed pinching antennas, which can be realized by applying small dielectric particles on a waveguide. Analytical results are first developed for the simple case with a single pinching antenna and a single waveguide, where the unique feature of the pinching-antenna system to create strong line-of-sight links and mitigate large-scale path loss is demonstrated. An advantageous feature of pinching-antenna systems is that multiple pinching antennas can be activated on a single waveguide at no extra cost; however, they must be fed with the same signal. This feature motivates the application of non-orthogonal multiple access (NOMA), and analytical results are provided to demonstrate the superior performance of NOMA-assisted pinching-antenna systems. Finally, the case with multiple pinching antennas and multiple waveguides is studied, which resembles a classical multiple-input single-input (MISO) interference channel. By exploiting the capability of pinching antennas to reconfigure the wireless channel, it is revealed that a performance upper bound on the interference channel becomes achievable, where the achievability conditions are also identified. Computer simulation results are presented to verify the developed analytical results and demonstrate the superior performance of pinching-antenna systems.    
\end{abstract}\vspace{-0.2em}

\vspace{-1.5em} 

\section{Introduction}
Recall that the data rate of a communication link impared by additive white Gaussian noise is bounded by Shannon capacity, $W\log_2 \left(1+\gamma_h\frac{ P_s}{P_n}\right)$, where $W$ denotes the available bandwidth, $\gamma_h$ denotes the effective channel gain, and $P_s$ and $P_n$ denote the signal and noise powers, respectively \cite{Cover1991}. Many recently developed communication techniques can be motivated by using the Shannon capacity formula \cite{you6g}. For example, the use of multiple-input multiple-output (MIMO) systems creates parallel channels between the transceivers, and increases the effective bandwidth to $N_{\rm MG}W$, where $N_{\rm MG}$ denotes the multiplexing gain and is related to the number of transceiver antennas \cite{Foschini,6375940}. Non-orthogonal multiple access (NOMA) is another example, which encourages spectrum sharing among multiple users and hence introduces extra degrees of freedom to configure $W$ and $P_s$ \cite{mojobabook,10729214}.  Noise modulation is another recently developed communication technique that treats $P_n$ as a configurable system parameter \cite{10373568}. Conventionally, a user's wireless channel, i.e.,  $\gamma_h$, has been viewed as a fixed system parameter that cannot be adjusted. Only recently, various flexible-antenna systems, such as reconfigurable intelligent surfaces (RISs) \cite{irs1,8741198}, intelligent reflecting surface (IRSs) \cite{irs2,9326394}, fluid-antenna systems \cite{9264694, 9650760}, and movable antennas \cite{10318061,10243545}, have been developed to make $\gamma_h$ also a reconfigurable system parameter. 

As the most well-known example of flexible-antenna systems, an RIS/IRS is equipped with a large number of low-cost reflecting elements, and deployed between transceivers \cite{irs1,8741198,irs2,9326394}. By intelligently adjusting the phase shifts of the reflecting elements, an RIS/IRS can dynamically reconfigure the transceivers' effective channel gains. As the latest members of the flexible-antenna system family, both fluid antennas and movable antennas are based on the idea to change the locations of the antennas at the transceivers, such that more favorable channel conditions are experienced by the transceivers \cite{9264694, 9650760, 10318061,10243545}. We note that for most existing flexible-antenna systems, their capabilities to combat large-scale path loss are limited. Take RIS/IRS as an example, where double attenuation can cause severe losses since the signal needs to go through both the transmitter-RIS/IRS link and the RIS/IRS-receiver link. Similarly, the current forms of fluid and movable antenna systems allow an antenna to be moved by at most a few wavelengths only, which has an insignificant impact on large-scale path loss. For example, if the line-of-sight (LoS) link between the transceivers is blocked, moving the antennas of the transceivers a few wavelengths is not helpful, particularly for high carrier frequencies (and hence small wavelengths). Furthermore, many existing flexible-antenna systems are expensive to build, where the flexibility to reconfigure the antennas, e.g., adding/removing antennas, is limited.

The aforementioned issues motivate the study of pinching antennas in this paper. The key idea of pinching antennas is illustrated in Fig. \ref{fig10x}, where pinching antennas are activated by applying small dielectric particles, e.g., plastic pinches, on a dielectric waveguide \cite{pinching_antenna2}. A demonstration carried out by DOCOMO in 2022 showed the following two unique features of pinching antennas \cite{pinching_antenna1}:\vspace{-0.5em}

\begin{itemize}[leftmargin=1em]
 \item Capability to support LoS communication: The use of pinching antennas can create a new LoS transceiver link or make an existing LoS link stronger, since the location of a pinching
antenna can be flexibly adjusted over a large scale and hence a pinching antenna can be easily deployed close to the target receiver to build a strong LoS link. 
\item Flexibility to reconfigure the antenna system: Increasing (or decreasing) the size of the pinching antenna system can be realized by simply applying additional pinches (or releasing existing ones).  Furthermore, multiple pinching antennas can be applied to one or multiple waveguides in a flexible and low-cost manner, which provides a new path forward for the implementation of MIMO \footnote{We note that pinching antennas can also be viewed as a type of leaky wave antennas, which have been used to design holographic  MIMO (H-MIMO) \cite{9848831}. However, the antenna spacing of H-MIMO is still at the wavelength scale, and hence, similar to the other flexible-antenna systems, its capability to combat large-scale path loss is also limited.\vspace{-0.5em}}. 
\end{itemize}
 The aim of this paper is two-fold. One is to develop practical designs of pinching-antenna systems, particularly for cases beyond a single pinching antenna, and the other is to provide a rigorous analysis of the performance achieved by pinching-antenna systems. In particular, this paper focuses on pinching-antenna assisted downlink transmission, and the contributions of the paper are listed as follows:\vspace{-0.5em}
\begin{itemize}[leftmargin=1em]
\item For the case with a single pinching antenna and a single waveguide, a closed-form expression for the ergodic sum rate achieved by the pinching-antenna system is developed. In addition, analytical results are also developed for the performance achieved by conventional antenna systems as a benchmark. The developed analytical results facilitate a performance comparison between systems employing conventional and pinching antennas, and illustrate the unique ability of pinching-antenna systems to create strong LoS links and mitigate large-scale path loss. Furthermore, the analysis shows that the performance gains of pinching antennas over conventional antennas are affected by the size of the area in which the users are deployed.  

 \begin{figure}[t]\centering \vspace{-2em}
    \epsfig{file=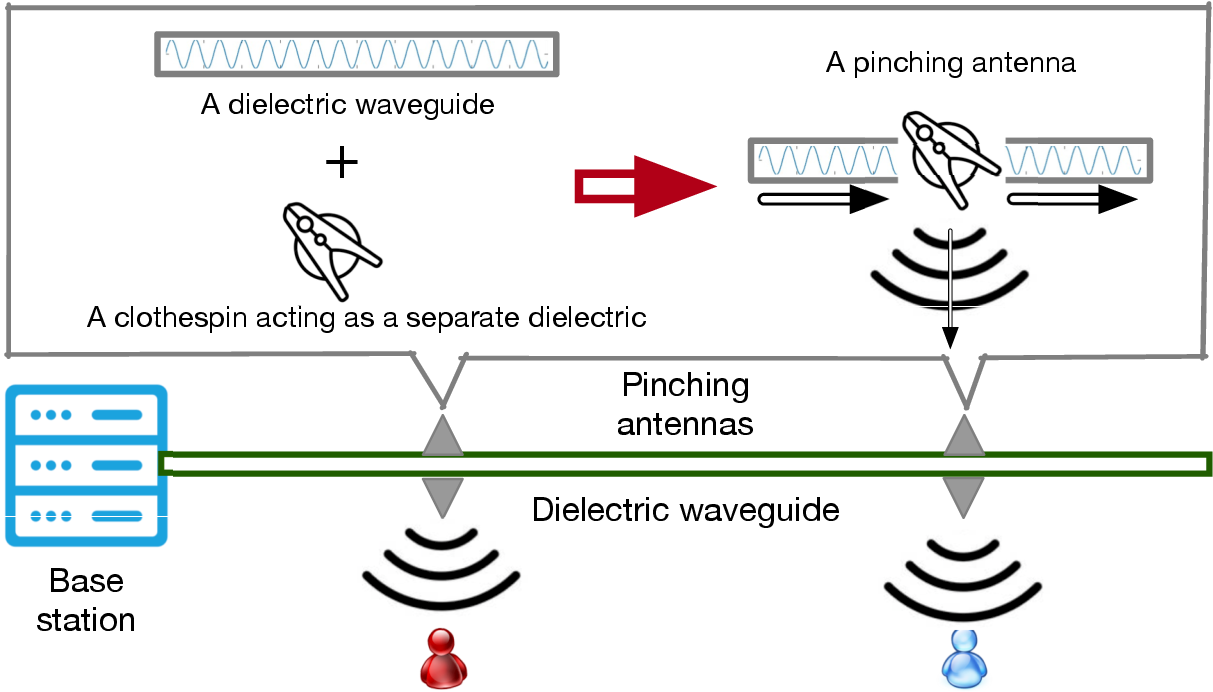, width=0.35\textwidth, clip=}\vspace{-0.5em}
\caption{Illustration of a pinching-antenna system \cite{pinching_antenna2}.  
  \vspace{-1em}    }\label{fig10x}   \vspace{-2.2em} 
\end{figure}

\item The fact that multiple pinching antennas can be activated at no extra cost motivates the study of systems with multiple pinching antennas and a single waveguide. We note that with more pinching antennas activated on a single waveguide, the transmit power of each antenna is reduced, which leads to the question of whether there is a benefit to using multiple pinching antennas. To obtain an insightful answer to this question, a time-division multiple access (TDMA)-assisted pinching-antenna system is considered first, where analytical results are developed to show that the users' data rates are monotonically increasing functions of the number of pinching antennas, i.e., the use of multiple pinching antennas is indeed beneficial.

\item How to use multiple pinching antennas on a single waveguide to serve multiple users simultaneously is also investigated. We note that if multiple pinching antennas are deployed on the same waveguide, they must be fed with the same signal, which is different from conventional MIMO systems. This observation means that a signal sent through the waveguide has to be a superimposed mixture of the signals of the multiple users to be served, which motivates the use of NOMA. In particular, by applying superposition coding at the base station and successive interference cancellation (SIC) at the users, multiple downlink users can be simultaneously served. Analytical results for the sum rates achieved by pinching antennas are derived and then used to demonstrate the superior performance of NOMA-assisted pinching-antenna systems, compared to those assisted by orthogonal multiple access (OMA). 

\item Multi-user multiple-input single-output (MISO) transmission can be supported by employing multiple waveguides and activating multiple pinching antennas on these waveguides. In this paper, the design of such pinching-antenna assisted MISO transmission is investigated, and its achievable performance is analyzed. In particular, the considered multi-user MISO scenario can be treated as a type of classical MISO interference channel \cite{Cover1991}, where there is a dilemma between using the principles of maximum-ratio combining (MRC) and zero-forcing (ZF) based beamforming. In particular, an MRC-based beamformer can maximize the strength of the intended signal, whereas a ZF-based beamformer can minimize the interference.  An ideal performance upper bound, i.e., a beamformer that can simultaneously boost the intended signal via MRC and suppress the interference via ZF, is generally not achievable for classical interference channels. However, in the context of pinching-antenna systems, the users' channels can be reconfigured by adjusting the locations of the antennas. Analytical results are presented to demonstrate that achieving the MISO upper bound is indeed possible with pinching antennas, where the achievability conditions are also identified. We note that these achievability results are also applicable to other types of flexible-antenna systems. 
 
\end{itemize} 

 
   \begin{figure}[t]\centering \vspace{-2em}
    \epsfig{file=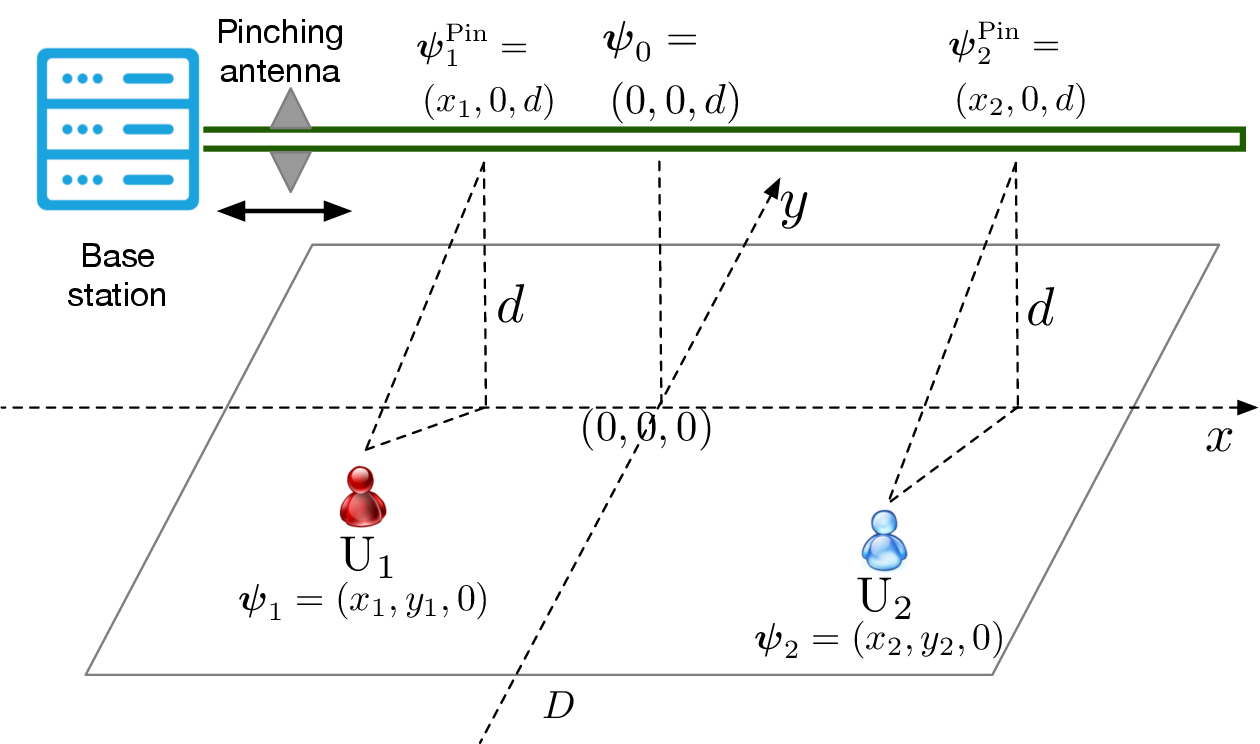, width=0.35\textwidth, clip=}\vspace{-0.5em}
\caption{Illustration of a network with a single waveguide and a single pinching antenna. In the time slot that serves ${\rm U}_m$, the pinching antenna at $\boldsymbol{\psi}^{\rm Pin}_m$ is activated.  
  \vspace{-1em}    }\label{fig1}   \vspace{-1em} 
\end{figure}

\section{Using A Single Pinching Antenna on A Single Waveguide}\label{section2}

Consider an OMA-based downlink communication scenario, where a base station serves $M$ single-antenna users, denoted by ${\rm U}_m$, $1\leq m \leq M$. Without loss of generality, TDMA is used as an example of OMA, i.e., ${\rm U}_m$ is served in time slot $m$.  It is assumed that the $M$ users are uniformly distributed in a square with side length $D$, where ${\rm U}_m$'s location is denoted by $\boldsymbol{\psi}_m$, as shown in Fig. \ref{fig1}. 

\vspace{-1em}   
\subsection{Conventional Antenna Systems}
A conventional antenna lacks installation flexibility and, hence, it has to be deployed at a fixed location. For the considered downlink scenario, a straightforward choice is to deploy the antenna at the center of the square, i.e., $\boldsymbol{\psi}_0$ shown in Fig. \ref{fig1}, where $d$ denotes the height of the antenna. Therefore, ${\rm U}_m$'s data rate is given by
\begin{align}
R_m^{\rm Conv}  =\frac{1}{M}\log_2\left(
1+\frac{\eta P_m}{ |\boldsymbol{\psi}_0 - \boldsymbol{\psi}_m| ^{ 2 }\sigma^2}
\right),
\end{align}
  where the factor $\frac{1}{M}$ is due to the use of TDMA, $\eta = \frac{c^2}{16\pi^2 f_c^2 }$, $c$ denotes the speed of light,  $f_c$ is the carrier frequency, $P_m$ denotes the transmit power for ${\rm U}_m$'s signal, $\sigma^2$ denotes the noise power, and $|\boldsymbol{\psi}_0 - \boldsymbol{\psi}_m| $ denotes the distance between the base station and ${\rm U}_m$. We note that for simplicity of illustration, it is assumed that there is an LoS link between each user and the base station, and hence, the free-space channel model is used, which can be justified by the typical application scenarios of pinching antennas, e.g., serving users/devices in lecture halls, shopping malls, factories \cite{pinching_antenna2, pinching_antenna1}. We note that the path loss exponents of non-line-of-sight (NLoS) links are larger than that of LoS links, which means that the performance gain of pinching antennas over conventional antennas in the NLoS case could be larger than those in the LoS case. The investigation of the performance of pinching antennas in the NLoS case is an important direction for future research but beyond the scope of this paper.

The ergodic sum rate achieved by conventional antenna systems is given by 
  \begin{align}
  R^{\rm Conv}_{\rm sum} = \sum^{M}_{m=1} \mathcal{E}_{\boldsymbol{\psi}_m}\left\{R_m^{\rm Conv}\right\} ,
  \end{align}  
  where $\mathcal{E}\{\cdot\}$ denotes expectation. 
  
  {\it Remark 1:} Due to the installation costs, the location of a conventional antenna has to be fixed. As a result, it is inevitable that some users will be far away from the base station, which will introduce excessive large-scale path loss and reduce the users' achievable data rates.  
  \vspace{-1em}

\subsection{Pinching-Antenna Systems}
The key feature of pinching antennas is their installation flexibility, i.e., they can be moved on a scale much larger than a wavelength and deployed right next to users. Throughout the paper, the following two notations are used for the locations of the pinching antennas:
\begin{itemize}
\item $\tilde{\boldsymbol \psi}_m^{\rm Pin}$: the {\bf general} notation for the location of the $m$-th pinching antenna;
\item $ {\boldsymbol \psi}_m^{\rm Pin}$: the {\bf specific}  location on the waveguide which is closest to ${\rm U}_m$, as shown in Fig. \ref{fig1}. 
\end{itemize}

 In this section, the case of using a single pinching antenna is considered.
During the $m$-th time slot, ${\rm U}_m$ is served, and the pinching antenna is moved to the location closest to the user\footnote{Similar to movable antennas, a pinching antenna is assumed to be movable on a pre-installed track parallel to the waveguide \cite{10318061,10243545}. Due to the low-cost feature of pinching antennas, a large number of pinching antennas can be pre-deployed on the track, which means that each pinching antenna is to cover only a small segment on the waveguide and hence can be moved to the required location on the waveguide in a short period of time. In this paper, we assume that a pinching antenna can be moved to a required location perfectly, where an important direction for future research is to study the impact of imprecise antenna positioning on the system performance.  }, e.g., ${\boldsymbol{\psi}}_{m}^{\rm Pin}$ shown in Fig. \ref{fig1}. Therefore,  ${\rm U}_m$'s achievable data rate can be expressed as follows\footnote{As this is an initial study of pinching antennas, the propagation loss in the waveguide is omitted, which makes our obtained results upper bounds on the performance achievable with pinching antennas. We note that the waveguide propagation loss is significantly smaller than the free-space path loss, e.g., the propagation loss of a dielectric waveguide at $28$ GHz is around $0.1$ dB/m, whereas the free-space path loss is around $40$ dB/m \cite{microwave}. }:
\begin{align}\label{1pa1wg}
  R^{\rm Pin}_m =\frac{1}{M}\log_2\left(
1+\frac{\eta P_m}{ |\boldsymbol{\psi}^{\rm Pin}_m - \boldsymbol{\psi}_m| ^{ 2 }\sigma^2}
\right),
\end{align} which means that the achievable sum rate is given by $
  R^{\rm Pin} = \sum^{M}_{m=1} R_m^{\rm Pin} $.

The ergodic sum rate achieved by the pinching antenna can be expressed as follows:
    \begin{align}
  R^{\rm Pin}_{\rm sum}  = &  \sum^{M}_{m=1} \mathcal{E}_{\boldsymbol{\psi}_m}\left\{\log_2\left(
1+\frac{\eta P_m}{ |\boldsymbol{\psi}^{\rm Pin}_m - \boldsymbol{\psi}_m| ^{ 2 }\sigma^2}
\right)\right\} .
  \end{align}
    To facilitate the performance analysis, the three-dimensional Cartesian coordinate system shown in Fig. \ref{fig1} is used, where the users are uniformly distributed within a square with its center at $(0,0,0)$. Furthermore, it is assumed that the square is in the $x$-$y$ plane, which means that ${\rm U}_m$'s location can be expressed as follows: ${\boldsymbol{\psi}}_{m}=(x_m, y_m, 0)$. The waveguide is assumed to be placed parallel to the $x$-axis, where the height of the waveguide is denoted by $d$. Therefore, the location of the antenna for the conventional antenna case is simply $\boldsymbol{\psi}_0=(0,0,d)$. During the $m$-th time slot, the location of the pinching antenna can be expressed as follows: ${\boldsymbol{\psi}}_{m}^{\rm Pin}=(x_m, 0, d)$.  

By using the above assumptions, the ergodic sum rate achieved by the pinching antenna is given by 
      \begin{align}
  R^{\rm Pin}_{\rm sum} =  & \frac{1}{M} \sum^{M}_{m=1} \mathcal{E}_{\boldsymbol{\psi}_m}\left\{\log_2\left(
1+\frac{\eta P_m}{ \left(
d^2+y_{ m}^2
\right)  \sigma^2}
\right)
 \right\}
 \\\nonumber
 =  & \frac{1}{M} \sum^{M}_{m=1} \int^{\frac{D}{2}}_{-\frac{D}{2}}\log_2\left(
1+\frac{\eta P_m}{ \left(
d^2+y_{ m}^2
\right) \sigma^2}
\right)
 \frac{1}{D}dy_m,
  \end{align}
  where the last step follows from the uniform deployment of the users. The expression for the ergodic sum rate can be further rewritten as follows: 
  \begin{align}\nonumber
  R^{\rm Pin}_{\rm sum} =  &\frac{1}{M}  \sum^{M}_{m=1}\int^{\frac{D}{2}}_{-\frac{D}{2}}\log_2\left(
 \frac{\left(
d^2+y_{ m}^2
\right)  +\frac{\eta P_m}{\sigma^2}}{ \left(
d^2+y_{ m}^2
\right)  }
\right)
 \frac{1}{D}dy_m  
 \\ \label{sum rate pinx3}
  =&\frac{2}{D}  \left(g\left(d^2+ \frac{\eta P_m}{\sigma^2} \right) - g\left(d^2 \right) \right),
  \end{align}
  where the last step is obtained by using the following definition: $g(a) \triangleq  \int^{\frac{D}{2}}_{0}\log_2\left(
 y ^2
  +a
\right)dy$. 
  
A closed-form expression for function $g(a)$ can be obtained as follows:  
   \begin{align}
  g(a) = & \int^{\frac{D}{2}}_{0}\log_2\left(
 y ^2+a 
\right) dy\\\nonumber = & \tau_2 -\log_2(e)D+2\log_2(e)\sqrt{a}\int^{\frac{D}{2\sqrt{a}}}_{0}  \frac{1}{ z ^2+1 }dz
  \\\nonumber
= & \tau_2 -\log_2(e)D+2\log_2(e)\sqrt{a} \tan^{-1}\left(\frac{D}{2\sqrt{a}}\right),
  \end{align}
  where $\tau_2=\frac{D}{2}\log_2\left(
 \frac{D^2}{4}+a
\right) $,  $z=\frac{y}{\sqrt{a}}$, and $\tan(\cdot)^{-1}$ denotes the inverse tangent function.

Therefore, the following lemma for the ergodic sum rate achieved by a pinching-antenna system can be obtained.

\begin{lemma}\label{lemma1}
The ergodic sum rate achieved by using a single pinching antenna on a single waveguide can be expressed as follows:
  \begin{align}\nonumber
  R^{\rm Pin}_{\rm sum} =  & \log_2\left(
 \frac{D^2}{4}+d^2+ \frac{\eta P_m}{\sigma^2} 
\right)\\\nonumber &+ \frac{4}{D} \log_2(e)\sqrt{d^2+\frac{\eta P_m}{\sigma^2}} \tan^{-1}\left(\frac{D}{2\sqrt{d^2+\frac{\eta P_m}{\sigma^2}}}\right)   \\\nonumber
  &- \log_2\left(
 \frac{D^2}{4}+d^2
\right)- \frac{4}{D}   \log_2(e)d\tan^{-1}\left(\frac{D}{2d}\right)  .
  \end{align}
  \end{lemma}
  To facilitate the performance comparison between conventional and pinching antenna systems, a high signal-to-noise ratio (SNR) approximation of $ R^{\rm Pin}_{\rm sum} $ is useful and can be obtained as follows. First, by applying a Taylor expansion,   a power series of  the inverse tangent function can be obtained as follows:
  \begin{align}
  \tan^{-1}\left(x\right) = \sum^{\infty}_{k=0}(-1)^k\frac{x^{2k+1}}{2k+1}. 
  \end{align}
  By using the series representation of $ \tan^{-1}\left(x\right)$, the term, $2\sqrt{d^2+\frac{\eta P_m}{\sigma^2}} \tan^{-1}\left(\frac{D}{2\sqrt{d^2+\frac{\eta P_m}{\sigma^2}}}\right)  $, can be   approximated  as follows:
    \begin{align}\nonumber
   &2\sqrt{d^2+\frac{\eta P_m}{\sigma^2}} \tan^{-1}\left(\frac{D}{2\sqrt{d^2+\frac{\eta P_m}{\sigma^2}}}\right) \\\nonumber =&
   \sum^{\infty}_{k=0}(-1)^k\frac{D^{2k+1} 2^{-k} \left(d^2+\frac{\eta P_m}{\sigma^2}\right)^{-k} }{2k+1} \approx
 D   ,
  \end{align}
  where the approximation is obtained by assuming $\frac{P_m}{\sigma^2}\rightarrow \infty$. Therefore, the following corollary for the high SNR approximation of $  R^{\rm Pin}_{\rm sum}$ can be obtained. 
  
  \begin{corollary}\label{corollary1}
  At high SNR, the ergodic sum rate of the pinching-antenna system can be approximated as follows:
  \begin{align}\label{approxm pinching an43}
  R^{\rm Pin}_{\rm sum} \approx & \log_2\left(
 \frac{D^2}{4}+d^2+ \frac{\eta P_m}{\sigma^2} 
\right)+ 2 \log_2(e)
    \\\nonumber
  &- \log_2\left(
 \frac{D^2}{4}+d^2
\right)- \frac{4}{D}   \log_2(e)d\tan^{-1}\left(\frac{D}{2d}\right)  .
  \end{align}
  \end{corollary}
  
By using Corollary \ref{corollary1}, the following conclusion regarding the performance difference between conventional and pinching-antenna systems can be obtained. 
\begin{lemma}\label{lemma2}
The sum rate achieved by a pinching-antenna system is always larger than that of a conventional antenna system, and the sum rate difference of the two systems, i.e., $ R^{\rm Pin}_{\rm sum}- R^{\rm Conv}_{\rm sum}$, is a monotonically increasing function of $\frac{D}{d}$ at high SNR. 
\end{lemma}
\begin{proof}
See Appendix \ref{proof-lemma2}. 
\end{proof}

  {\it Remark 1:} The performance gain shown in Lemma \ref{lemma2} is due to the capability of pinching antennas to create strong LoS links and reduce large-scale path loss. We note that the use of pinching antennas can also reduce the blockage of LoS links. Recall that the probability of having an LoS link is a function of the transceiver distance, i.e.,  $\mathbb{P}({\rm LoS}) = e^{-\lambda_{\rm LoS}r}$, where $\lambda_{\rm LoS}$ is a system parameter related to the building density and $r$ denotes the transceiver distance  \cite{5621983,7448962}. Since the use of pinching antennas reduces $r$, the LoS blockage probability can be reduced by pinching antennas. Therefore, an important direction for future research is to study the performance of pinching antennas in the presence of LoS blockages.  
  
  {\it Remark 2:} The analysis carried out in this section assumes that the users are located within a square. We note that the shape of the area in which the users are deployed has a significant impact on the performance gain of pinching antennas over conventional antennas. For example, simulation results will be provided in Section \ref{section5} to show that the performance gain of pinching antennas increases significantly if the users are deployed in a rectangular area.

  \section{Using Multiple Pinching Antennas on A Single Waveguide }\label{section3}
Without loss of generality, assume that    $N$ pinching antennas are activated on a single waveguide, where the location of the $n$-th pinching antenna is denoted by $\tilde{\boldsymbol \psi}_n^{\rm Pin}$, $1\leq n \leq N$\footnote{If $\tilde{{\boldsymbol \psi}}_n^{\rm Pin}= {{\boldsymbol \psi}}_m^{\rm Pin}$, the $n$-th pinching antenna is deployed at the point on the waveguide which is closest to ${\rm U}_m$. An illustration of ${{\boldsymbol \psi}}_m^{\rm Pin}$ is shown in Fig. \ref{fig1}. }. Because the base station is equipped with multiple antennas, it is natural to serve $M$ users simultaneously. Collect the signals sent by the $N$ pinching antennas in a vector denoted by $\mathbf{s}$. By treating the $N$ pinching antennas as conventional linear array antennas, the received signal at  ${\rm U}_m$ can be expressed as follows:
 \begin{align}
 \label{miso model1}
  y_m = \mathbf{h}_m^H\mathbf{s}+w_m,
  \end{align}
  where  $w_m$ denotes the additive white Gaussian noise,  
  \begin{align}
  \mathbf{h}_m =  \begin{bmatrix} 
 \frac{\eta^{\frac{1}{2}} e^{-j\frac{2\pi }{\lambda}\left| {\boldsymbol \psi}_m  -\tilde{\boldsymbol \psi}_1^{\rm Pin}\right|}}{\left| {\boldsymbol \psi} _m -\tilde{{\boldsymbol \psi}}_1^{\rm Pin}\right|} &\cdots & \frac{\eta^{\frac{1}{2}}  e^{-j\frac{2\pi }{\lambda}\left| {\boldsymbol \psi}_m  -\tilde{{\boldsymbol \psi}}_N^{\rm Pin}\right|}}{  \left| {\boldsymbol \psi} _m -\tilde{\boldsymbol \psi}_N^{\rm Pin}\right|} 
 \end{bmatrix}^T,
 \end{align}
 $\lambda=\frac{2\pi}{f_c}$, and the spherical wave channel model is used \cite{9738442}.

 The system model shown in \eqref{miso model1} suggests that conventional MISO transmission strategies can be straightforwardly applied in the pinching-antenna system, which, however, is not true, as explained in the following. Recall that the $N$ pinching antennas are located on the same waveguide, which means that the signal sent by one pinching antenna is a phase-shifted version of the signal sent by another pinching antenna \cite{microwave}. Therefore, the signal vector $\mathbf{s}$ can be expressed as follows:
 \begin{align}\label{miso model2}
 \mathbf{s} =\sqrt{\frac{P}{N}}\begin{bmatrix} e^{-j\theta_1} &\cdots &e^{-j\theta_N} \end{bmatrix}^T s,
 \end{align}
 where $\theta_n$ denotes the phase shift experienced at the $n$-th antenna, $P$ denotes the total transmit power, and $s$ is the signal passed onto the waveguide. We note that $\theta_n$ is the phase shift for a signal traveling from the feed point of the waveguide to the $n$-th pinching antenna, and hence is a function of the location of this antenna, e.g., $\theta_n =2\pi \frac{\left| {\boldsymbol \psi}_0^{\rm Pin}  -\tilde{\boldsymbol \psi}_n^{\rm Pin}\right|}{\lambda_g} $, where ${\boldsymbol \psi}_0^{\rm Pin}$ denotes the location of the feed point of the waveguide, and $\lambda_g$ denotes the waveguide wavelength in a dielectric waveguide. We further note that $ {\lambda}_g=\frac{\lambda}{n_{\rm eff}}$, where $n_{\rm eff}$ denotes the effective refractive index of a dielectric waveguide \cite{microwave}. To facilitate insightful performance analysis, we assume that the overall transmit power can be equally shared among the $N$ activated pinching antennas. The use of practical waveguide specifications, including propagation losses, to characterize the transmit power allocation among the multiple pinching antenna is an important direction for future research.

By combining \eqref{miso model1} and \eqref{miso model2}, the signal received by ${\rm U}_m$ can be expressed as follows:
    \begin{align} \label{miso model3}
  y_m =\left(\sum^{N}_{n=1}\frac{\eta^{\frac{1}{2}}  e^{-j\frac{2\pi }{\lambda}\left| {\boldsymbol \psi}_m  -\tilde{\boldsymbol \psi}_n^{\rm Pin}\right|}}{  \left| {\boldsymbol \psi} _m -\tilde{\boldsymbol \psi}_n^{\rm Pin}\right|} e^{-j\theta_n} \right)\sqrt{\frac{P}{N}} s+w_m.
  \end{align}

The system model in \eqref{miso model3} shows a unique feature of pinching antennas. Compared to the use of conventional antennas, the use of pinching antennas offers more degrees of freedom, since both the large-scale path loss, $\left| {\boldsymbol \psi}_m  -\tilde{\boldsymbol \psi}_n^{\rm Pin}\right|$, and the phase shifts, $\theta_n$, can be reconfigured by positioning the pinching antennas. Compared to the other types of flexible antennas,  the pinching-antenna system has fewer degrees of freedom. Take movable antennas as an example, where different movable antennas can be fed with independent signals. These features of the pinching-antenna system can be illustrated better by the following two special cases. 
\vspace{-1em}
\subsection{OMA-Assisted Pinching-Antenna Systems}\label{subsection NP}
The system model in \eqref{miso model3} is similar to conventional hybrid beamforming with a single radio-frequency (RF) chain \cite{7434598}. Given a single RF chain, it is natural to consider the case in which the users are served individually, i.e., only ${\rm U}_m$ is served in time slot $m$.  In this case,  ${\rm U}_m$'s  data rate achieved by multiple pinching antennas is given by
    \begin{align} \label{miso model3xx}
  R_m =\frac{1}{M}\log\left(1+\left|\sum^{N}_{n=1}\frac{\eta^{\frac{1}{2}}  e^{-j\frac{2\pi }{\lambda}\left| {\boldsymbol \psi}_m  -\tilde{\boldsymbol \psi}_n^{\rm Pin}\right|}}{  \left| {\boldsymbol \psi} _m -\tilde{\boldsymbol \psi}_n^{\rm Pin}\right|} e^{-j\theta_n}\right|^2  {\frac{P_m}{N\sigma^2}} \right) ,
  \end{align}
  where   $P_m$ denotes the transmit power for ${\rm U}_m$. Assume that the location of each pinching antenna can be finely tuned such that $\frac{2\pi }{\lambda}\left| {\boldsymbol \psi}_m  -\tilde{\boldsymbol \psi}_n^{\rm Pin}\right|+\theta_n=2k\pi$, where $k$ is an arbitrary integer. By using this assumption, an upper bound on ${\rm U}_m$'s data rate can be obtained as follows: 
  \begin{align}\label{upper boundx3}
  R_m \leq  &\frac{1}{M}\log_2\left(1+\frac{P_m}{N\sigma^2}\left(\sum^{N}_{n=1}\frac{\eta^{\frac{1}{2}}  }{  \left| {\boldsymbol \psi} _m -\tilde{\boldsymbol \psi}_n^{\rm Pin}\right|}  \right)^2\right).
  \end{align}

We recall that when ${\rm U}_m$ is served, it is ideal to place all pinching antennas as close to ${\boldsymbol \psi}_m^{\rm Pin}$ as possible, since ${\boldsymbol \psi}_m^{\rm Pin}$ is the location on the waveguide closest to ${\rm U}_m$, as shown in Fig. \ref{fig1}. Moving an antenna a few wavelengths for satisfying  $\frac{2\pi }{\lambda}\left| {\boldsymbol \psi}_m  -\tilde{\boldsymbol \psi}_n^{\rm Pin}\right|+\theta_n=2k\pi$ has a limited impact on the distance $ \left| {\boldsymbol \psi} _m -\tilde{\boldsymbol \psi}_n^{\rm Pin}\right|$. The above discussions justify the following assumption: $\frac{\left| {\boldsymbol \psi} _m -\tilde{\boldsymbol \psi}_n^{\rm Pin}\right|}{\left| {\boldsymbol \psi} _m - {\boldsymbol \psi}_m^{\rm Pin}\right|}\approx 1$, i.e.,  the $N$ pinching antennas are clustering around ${\boldsymbol \psi}_m^{\rm Pin}$.  

If the assumption that $\frac{\left| {\boldsymbol \psi} _m -\tilde{\boldsymbol \psi}_n^{\rm Pin}\right|}{\left| {\boldsymbol \psi} _m - {\boldsymbol \psi}_m^{\rm Pin}\right|}\approx 1$ is feasible, the upper bound  on $R_m$ shown in \eqref{upper boundx3} can be simplified as follows:
  \begin{align}\nonumber
  R_m \leq &  
  \frac{1}{M}   \log_2\left(1+\frac{P_m}{N\sigma^2}\left(\sum^{N}_{n=1}\frac{\eta^{\frac{1}{2}}  }{  \left| {\boldsymbol \psi} _m - {\boldsymbol \psi}_m^{\rm Pin}\right|}  \right)^2\right)\\\label{mulpa1wg}=& \frac{1}{M} \log_2\left(1+\frac{N P_m\eta}{ \sigma^2} \frac{1  }{  \left| {\boldsymbol \psi} _m -{\boldsymbol \psi}_m^{\rm Pin}\right|^2}   \right).
  \end{align}  
 Compared to the data rate shown in \eqref{1pa1wg},   \eqref{mulpa1wg} shows that the use of multiple pinching antennas can significantly improve the performance of the pinching-antenna system. 
 
The upper bound on $R_m$ shown in \eqref{mulpa1wg} can be realized by the following location search algorithm.

\begin{itemize}
\item  The location of the first pinching antenna is obtained by focusing on the segment between $ {\boldsymbol \psi}_m^{\rm Pin} $ and the end of the waveguide and using the first-found location which satisfies ${\rm mod}\left\{ \frac{2\pi }{\lambda}\left| {\boldsymbol \psi}_m  -\tilde{\boldsymbol \psi}_1^{\rm Pin}\right|+\theta_1,  2\pi\right\}=0$, where ${\rm mod}\{a,b\}$ denotes the modulo operation of $a$ by $b$. 

\item Sucessively, the location of the $n$-th pinching antenna can be obtained by focusing on the segment between $\tilde{\boldsymbol \psi}_{n-1}^{\rm Pin}+\tilde{\Delta}$ and the end of the waveguide and using the first-found location which satisfies ${\rm mod}\left\{ \frac{2\pi }{\lambda}\left| {\boldsymbol \psi}_m  -\tilde{\boldsymbol \psi}_n^{\rm Pin}\right|+\theta_n, 2\pi\right\}=0$, where $\tilde{\Delta}$ is the guard distance to avoid antenna coupling.

 \end{itemize}
 
 {\it Remark 3:} We note that the upper bound shown in \eqref{mulpa1wg} is not achievable by other types of antenna systems, since a conventional antenna needs to be installed at a fixed location and other flexible antennas can be moved by a few wavelengths only (i.e., they cannot be moved from ${\boldsymbol \psi}_m^{\rm Pin} $ in time slot $m$ to ${\boldsymbol \psi}_n^{\rm Pin} $ in time slot $n$). Furthermore, the use of pinching antennas reduces the hardware cost since only one RF chain is needed, and also yields a flexible antenna configuration since adding/removing pinching antennas incurs almost no additional costs \cite{pinching_antenna2}. 
 
  \begin{figure}[t]\centering \vspace{-0.2em}
    \epsfig{file=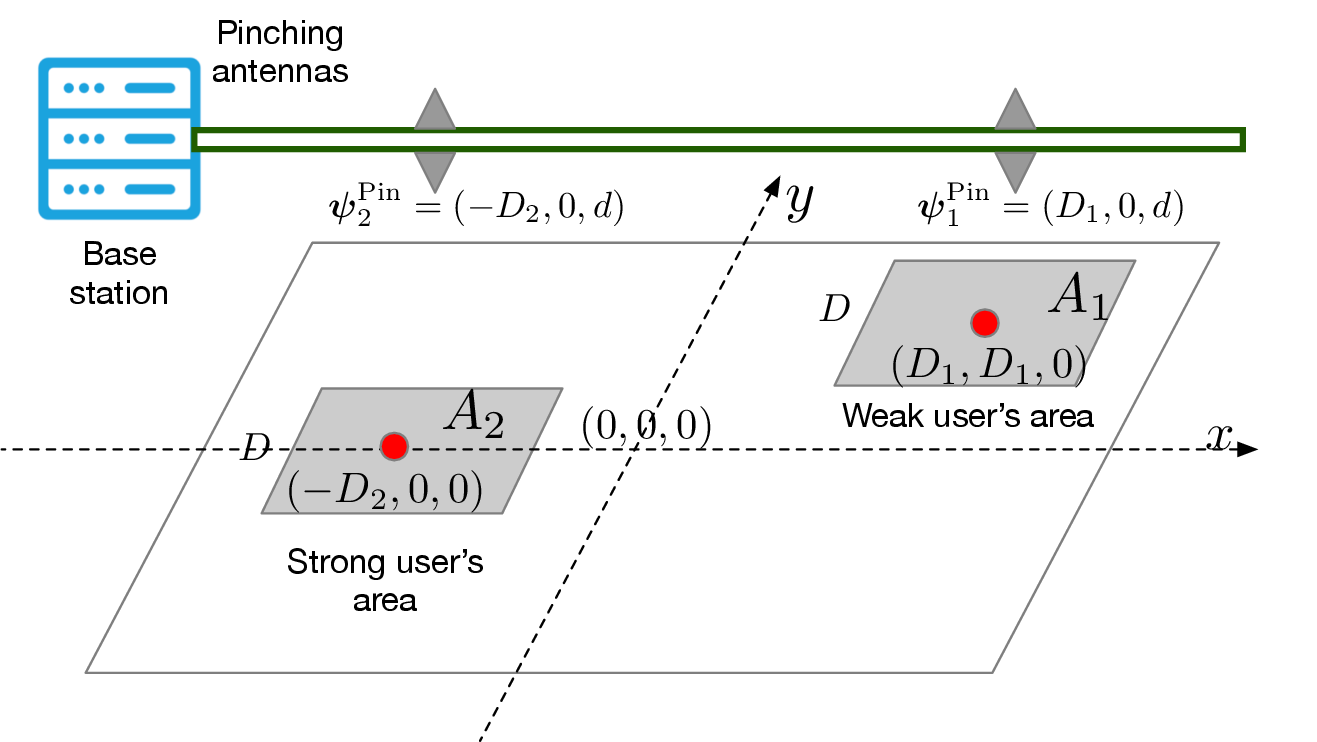, width=0.35\textwidth, clip=}\vspace{-0.5em}
\caption{Illustration of a NOMA-assisted pinching-antenna system, with a single waveguide and two pinching antennas. The weak user is uniformly deployed in the square denoted by  $A_1$ with its center at $(D_1,D_1,0)$, and the strong user is uniformly deployed in the square denoted by  $A_2$ with its center at $(-D_2,0,0)$. The side lengths of the two squares are identical and denoted by $D$. 
  \vspace{-1em}    }\label{fig2}   \vspace{-0.5em} 
\end{figure}
\vspace{-1em}  
 \subsection{NOMA Assisted Pinching-Antenna Systems}\label{subsection NOMA}
For simplicity, we focus on the case of $M=N$. If multiple users are to be served simultaneously, the users' signals need to be superimposed, i.e., $s$ in \eqref{miso model2} is a mixture of multiple users' signals, which motivates the application of NOMA. In particular, consider $s=\sum^{M}_{m=1}\sqrt{\alpha}_ms_m$, where $s_m$ denotes ${\rm U}_m$'s signal,   $\alpha_m$ denotes the power allocation coefficient for ${\rm U}_m$, and $\sum^{M}_{m=1} {\alpha}_m=1$.  Assume that the $n$-th pinching antenna is set as $\tilde{\boldsymbol \psi}_n^{\rm Pin}={\boldsymbol \psi}_n^{\rm Pin}$, i.e., the $n$-th pinching antenna is placed to be closest to ${\rm U}_n$,  and the users are ordered according to their channel conditions in an ascending order, i.e., $|h_{1}|^2\leq \cdots \leq |h_{M}|^2$, where $h_i=\sum^{M}_{n=1}\frac{\eta^{\frac{1}{2}}  e^{-j\frac{2\pi }{\lambda}\left| {\boldsymbol \psi}_i  -\tilde{\boldsymbol \psi}_n^{\rm Pin}\right|}}{  \left| {\boldsymbol \psi} _i -\tilde{\boldsymbol \psi}_n^{\rm Pin}\right|} e^{-j\theta_n} $. How this channel order can be realized will be discussed later.  According to the principle of power-domain NOMA, ${\rm U}_m$ will decode ${\rm U}_j$'s signal, $1\leq j \leq m-1$, before decoding its own signal, which means that the data rate of ${\rm U}_m$'s signal is given by \cite{Nomading}
 \begin{align}\label{um1 rate}
 R_m =   \min \left\{R_{m,m}, \cdots, R_{M,m} \right\},
 \end{align}
 where $R_{i,m}$ denotes the data rate for ${\rm U}_i$ to decode ${\rm U}_m$'s signal, i.e., $R_{i,m}= \log\left(1+ \frac{|h_i|^2 \frac{P}{M}\alpha_m}{\sum^{M}_{j=m+1}|h_i|^2\frac{ P}{M} \alpha_j+\sigma^2} \right)$ for $i\geq m$. We note that $\frac{P}{M}$ is used in $R_{i,m}$ since $M$ pinching antennas are activated and $P$ is equally shared among the antennas\footnote{We note that the power allocation coefficients, $\alpha_m$, ensure that the power of a signal passed within the waveguide (i.e., $s$) is $P$, and the power of the signal sent by each of the $M$ pinching antenna is $\frac{P}{M}$.    }. 
 
 \subsubsection{User Scheduling} Recall that the users' channel gains, $h_m=\sum^{M}_{n=1}\frac{\eta^{\frac{1}{2}}  e^{-j\frac{2\pi }{\lambda}\left| {\boldsymbol \psi}_m  -\tilde{\boldsymbol \psi}_n^{\rm Pin}\right|}}{  \left| {\boldsymbol \psi} _m -\tilde{\boldsymbol \psi}_n^{\rm Pin}\right|} e^{-j\theta_n} $, contain sums of complex numbers. To ensure constructive superposition, a sophisticated optimization of the antenna locations is required, which can cause high computational complexity. A low-complexity alternative is to apply user scheduling. In particular, users that are far away from each other are scheduled for the implementation of NOMA, which also justifies the choice of $ \tilde{\boldsymbol \psi}_m^{\rm Pin}={\boldsymbol \psi}_m^{\rm Pin}$. Because of the large-scale path loss, the proposed scheduling can ensure that the term $ \left| {\boldsymbol \psi} _m-{\boldsymbol \psi}_m^{\rm Pin}\right|$ is dominant compared to the other terms in $h_m$, an effect similar to the frequency reuse in cellular networks.  As a result, there is no need for fine-tuned antenna placement.  The details will be illustrated for the special case of $M=2$ in the following. 
 
 
 \subsubsection{The Case of $M=N=2$} 
As shown in Fig. \ref{fig2}, the scheduled weak user (${\rm U}_1$) is uniformly distributed in $A_1$, i.e., a square with side length $D$ and its center at $(D_{1},D_1,0)$, and the scheduled strong user (${\rm U}_2$) is uniformly distributed in $A_2$, another square with side length $D$ and its center at $(D_{2},0,0)$. As long as the two areas, $A_1$ and $A_2$, are far away from each other, i.e., $D_1$ is large, $h_m$ can be simplified as follows: $|h_m|^2\approx   \frac{\eta  }{  \left| {\boldsymbol \psi} _m -{\boldsymbol \psi}_m^{\rm Pin}\right|^2}   $, $m\in \{1,2\}$. Furthermore, assume that $D_1\geq D$ which  guarantees the assumption $|h_1|^2\leq |h_2|^2$. With the simplified expressions of the channel gains, the two users'  data rates shown in \eqref{um1 rate} can be simplified as follows:
 \begin{align}\label{um1 rate2}
 R_1 \approx  \log_2  \left(1+ \frac{\frac{\eta  }{  \left| {\boldsymbol \psi} _1 -{\boldsymbol \psi}_1^{\rm Pin}\right|^2}  \frac{P}{N}\alpha_1}{ \frac{\eta  }{  \left| {\boldsymbol \psi} _1 -{\boldsymbol \psi}_1^{\rm Pin}\right|^2}   \frac{ P}{N} \alpha_2+\sigma^2} \right),
 \end{align}
 and 
  \begin{align}\label{um1 rate3}
 R_2 \approx  \log_2   \left(1+  {\frac{\eta  }{  \left| {\boldsymbol \psi} _2 -{\boldsymbol \psi}_2^{\rm Pin}\right|^2}  \frac{P}{N\sigma^2}\alpha_2}\right). 
 \end{align}

By exploiting the uniform distribution of the users' locations, ${\rm U}_2$'s ergodic data rate can be evaluated as follows: 
    \begin{align}
 \mathcal{E}_{\boldsymbol{\psi}_2}\left\{R_2
 \right\}  = &  \mathcal{E}_{\boldsymbol{\psi}_2}\left\{\log_2\left(1+  {\frac{\eta  }{  \left| {\boldsymbol \psi} _2 -{\boldsymbol \psi}_2^{\rm Pin}\right|^2}  \frac{P}{N\sigma^2}\alpha_2}\right)
 \right\}\\\nonumber 
 =&\frac{1}{D}
\int^{\frac{D}{2}}_{-\frac{D}{2}}\log_2\frac{  y_2^2+d^2+  {  \frac{\eta P}{N \sigma^2}\alpha_2} }{y_2^2+d^2 }
dy_2  . 
 \end{align} 
 By using the function $g(a)$ developed in the previous section, ${\rm U}_2$'s ergodic data rate can be expressed  as follows: 
     \begin{align}
 \mathcal{E}_{\boldsymbol{\psi}_2}\left\{R_2
 \right\}  = &  \frac{2}{D}g\left(
 d^2+  {  \frac{\eta P}{N \sigma^2}\alpha_2}
 \right)   -\frac{2}{D} g\left(
 d^2
 \right) .
 \end{align} 
  Following steps similar to those in the previous section, ${\rm U}_2$'s ergodic data rate can be approximated at high SNR as follows: 
   \begin{align}\label{r22}
 \mathcal{E}_{\boldsymbol{\psi}_1}\left\{R_2
 \right\}  \approx & \log_2\left(
 \frac{D^2}{4}+d^2+ \frac{\eta P\alpha_2}{N\sigma^2} 
\right)+ 2 \log_2(e)
    \\\nonumber
  &- \log_2\left(
 \frac{D^2}{4}+d^2
\right)- \frac{4}{D}   \log_2(e)d\tan^{-1}\left(\frac{D}{2d}\right)  .
  \end{align} 
 Similarly, ${\rm U}_1$'s   ergodic data rate, $ \mathcal{E}_{\boldsymbol{\psi}_1}\left\{R_1
 \right\} $,  can be expressed as follows:
   \begin{align}
 \mathcal{E}_{\boldsymbol{\psi}_1}\left\{R_1
 \right\}  
 = &
 \mathcal{E}_{\boldsymbol{\psi}_1}\left\{\log_2
  \left(1+ \frac{\frac{\eta  }{  \left| {\boldsymbol \psi} _1 -{\boldsymbol \psi}_1^{\rm Pin}\right|^2}  \frac{P}{N}\alpha_1}{ \frac{\eta  }{  \left| {\boldsymbol \psi} _1 -{\boldsymbol \psi}_1^{\rm Pin}\right|^2}   \frac{ P}{N} \alpha_2+\sigma^2} \right)
 \right\}  \\\nonumber     = &\frac{1}{D}
\int_{D_1-\frac{D}{2}}^{D_1+\frac{D}{2}}\log_2
  \left(     y_1^2+d^2 +    \frac{\eta P}{N\sigma^2}  \right)
dy_1\\\nonumber 
&-\frac{1}{D}
\int_{D_1-\frac{D}{2}}^{D_1+\frac{D}{2}}\log_2
  \left(   y_1^2+    \frac{ \eta P}{N\sigma^2} \alpha_2+d^2 \right)
dy_1.
 \end{align}
 It is challenging to find a closed-form expression for $ \mathcal{E}_{\boldsymbol{\psi}_1}\left\{R_1
 \right\} $. Therefore, a high-SNR asymptotic study will be carried out in the following. In particular, at high SNR, ${\rm U}_1$'s   ergodic data rate, $ \mathcal{E}_{\boldsymbol{\psi}_1}\left\{R_1
 \right\} $,  can be approximated by the following constant:  
   \begin{align}\label{r11}
 \mathcal{E}_{\boldsymbol{\psi}_1}\left\{R_1
 \right\}  
 \approx  &
 \mathcal{E}_{\boldsymbol{\psi}_1}\left\{\log_2
  \left(1+ \frac{\alpha_1}{\alpha_2}\right)
 \right\}    =-\log_2 \alpha_2,
 \end{align} 
 where the fact that $\alpha_1+\alpha_2=1$ is used. By combining \eqref{r22} with \eqref{r11}, the following lemma can be obtained.
 
 \begin{lemma}\label{lemma3} 
For large $D_1$ and at high SNR, a closed-form approximation for the ergodic sum rate achieved by a NOMA-assisted pinching-antenna system can be obtained as follows:  
 \begin{align}\nonumber
 &R_{\rm sum}^{\rm NOMA} \approx -\log_2 \alpha_2+  \log_2\left(
 \frac{D^2}{4}+d^2+ \frac{\eta P\alpha_2}{N\sigma^2} 
\right)+ 2 \log_2(e)
    \\ \label{lemma3eq}
  &- \log_2\left(
 \frac{D^2}{4}+d^2
\right)- \frac{4}{D}   \log_2(e)d\tan^{-1}\left(\frac{D}{2d}\right) .
 \end{align}
 \end{lemma}
An interesting question is how the OMA-assisted system discussed in Section \ref{subsection NP} compares to the NOMA one proposed in this section, which motivates the following asymptotic study. At high SNR and for large $D_1$, the instantaneous sum rate for the NOMA system can be approximated as follows: 
  \begin{align}\nonumber
 R_{\rm sum} ^{\rm NOMA}\approx & -\log_2 \alpha_2+\log_2\left(  {\frac{\eta  }{  \left| {\boldsymbol \psi} _2 -{\boldsymbol \psi}_2^{\rm Pin}\right|^2}  \frac{P}{N\sigma^2}\alpha_2}\right)
 \\\nonumber =  &  -\log_2\left(  {  \left| {\boldsymbol \psi} _2 -{\boldsymbol \psi}_2^{\rm Pin}\right|^2}   \right)+\log_2\left(   \frac{\eta P}{\sigma^2} \right)-\log_2 N. 
 \end{align}
 On the other hand, recall from \eqref{mulpa1wg} that the two users' data rates for OMA are $ \frac{1}{M} \log_2\left(1+\frac{N P_m\eta}{ \sigma^2} \frac{1  }{  \left| {\boldsymbol \psi} _m -{\boldsymbol \psi}_m^{\rm Pin}\right|^2}   \right)$, which means that the high-SNR approximation of
 the instantaneous sum rate for the OMA case is given by
  \begin{align}\nonumber
 R_{\rm sum}^{\rm OMA}\approx &   \log_2\left( {  \frac{\eta P_m}{\sigma^2} }\right)- \frac{1}{2} \log_2\left( { {  \left| {\boldsymbol \psi} _2 -{\boldsymbol \psi}_2^{\rm Pin}\right|^2}  }\right)\\\nonumber &-\frac{1}{2} \log_2\left(   { {  \left| {\boldsymbol \psi} _1 -{\boldsymbol \psi}_1^{\rm Pin}\right|^2}    }\right) +\log_2 N.
 \end{align}
Recall that for NOMA, the total transmit power for the $M$ users over $M$ time slots is $P$, whereas for OMA, $P_m$ denotes each user's transmit power over one time slot. For a fair comparison, $MP=P_m$ is assumed. 
 Therefore, for the case of $M=N=2$, the difference between the two sum rates is given by
   \begin{align}\nonumber
& R_{\rm sum}^{\rm NOMA}- R_{\rm sum}^{\rm OMA}\\\nonumber \approx  &   -\frac{1}{2} \log_2\left( { {  \left| {\boldsymbol \psi} _2 -{\boldsymbol \psi}_2^{\rm Pin}\right|^2}  }\right)+\frac{1}{2} \log_2\left(   { {  \left| {\boldsymbol \psi} _1 -{\boldsymbol \psi}_1^{\rm Pin}\right|^2}    }\right) -3 
  \\\label{gapgain} 
 =  &   \log_2\left( {\frac { \left| {\boldsymbol \psi} _1 -{\boldsymbol \psi}_1^{\rm Pin}\right| }{  \left| {\boldsymbol \psi} _2 -{\boldsymbol \psi}_2^{\rm Pin}\right|}  }\right)-3,
 \end{align}
 which is guaranteed to be positive for the case of large $D_1$, i.e., $\left| {\boldsymbol \psi} _1 -{\boldsymbol \psi}_1^{\rm Pin}\right| \gg\left| {\boldsymbol \psi} _2 -{\boldsymbol \psi}_2^{\rm Pin}\right| $. 
 
  
   \begin{figure}[t]\centering \vspace{-0.2em}
    \epsfig{file=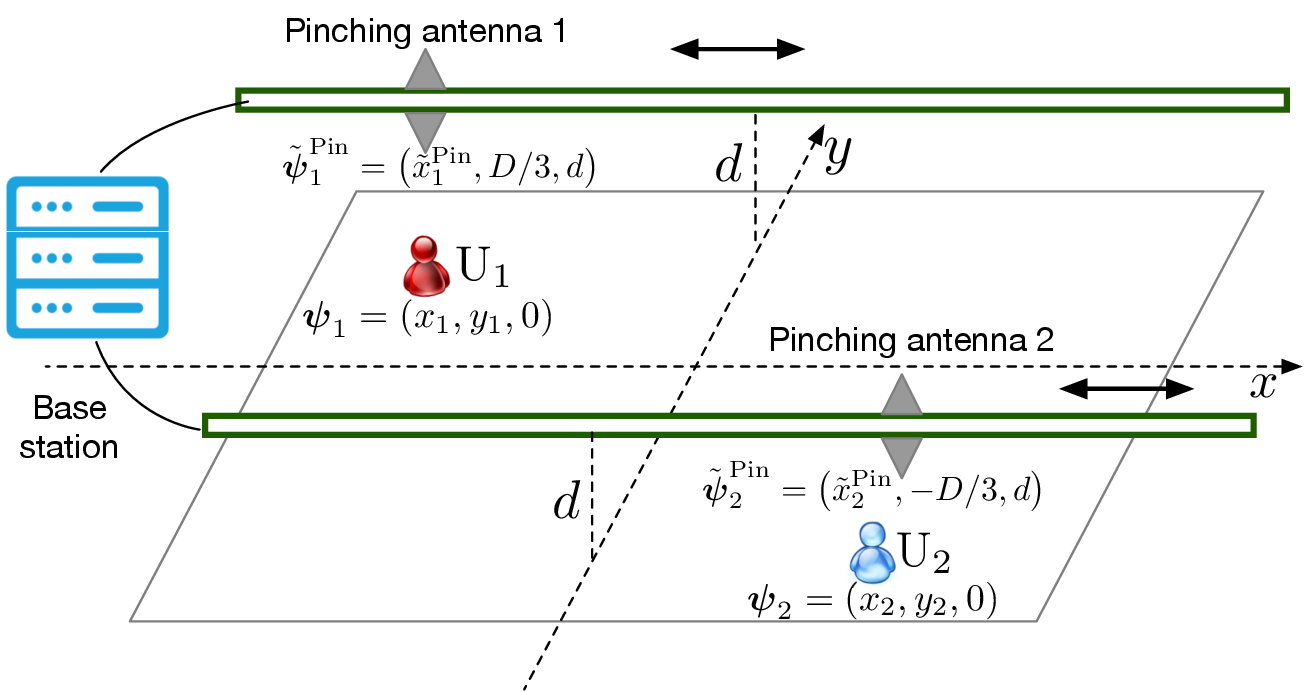, width=0.35\textwidth, clip=}\vspace{-0.5em}
\caption{Illustration of a network with multiple waveguides and multiple pinching antennas.   
  \vspace{-1em}    }\label{fig3}   \vspace{-0.5em} 
\end{figure}
  \section{Using Multiple Pinching Antennas on Multiple Waveguides  }
  \label{section4}
This section focuses on the use of $K$ waveguides, where a single pinching antenna is activated on each waveguide. Denote the location of the pinching antenna on the $k$-th waveguide by $\tilde{\boldsymbol \psi}_{k}^{\rm Pin}$.  
We note that different waveguides can be fed with   different signals, which means that the received signal at ${\rm U}_m$ can be expressed as follows:   
   \begin{align}\nonumber
  v_m =& \sum^{K}_{k=1}h_{m,k} p_{m,k} \sqrt{P}s_m +\sum_{i\neq m}\sum^{K}_{k=1}h_{m,k}  p_{i,k} \sqrt{P}s_i +w_m,
  \end{align}
 where the channel between ${\rm U}_m$ and the $k$-th antenna  is denoted by $h_{m,k}=\frac{\sqrt{\eta} e^{-2\pi j \left(\frac{  1}{\lambda}\left| {\boldsymbol \psi}_m  -\tilde{\boldsymbol \psi}_k^{\rm Pin}\right|
  +\frac{1}{\lambda_g}\left| {\boldsymbol \psi}_0^{\rm Pin}  -\tilde{\boldsymbol \psi}_k^{\rm Pin}\right|
  \right)}}{  \left| {\boldsymbol \psi} _m -\tilde{\boldsymbol \psi}_k^{\rm Pin}\right|} $, $P$ denotes the overall transmit power for all users, $p_{m,k}$ is the beamforming coefficient assigned to ${\rm U}_m$'s signal on the $k$-th waveguide. We note that the phase shifts    $e^{-2\pi j  \frac{1}{\lambda_g}\left| {\boldsymbol \psi}_0^{\rm Pin}  -\tilde{\boldsymbol \psi}_k^{\rm Pin}\right|
  }$ are due to the signals' propagation through the waveguide, and the phase shifts $e^{-2\pi j  \frac{  1}{\lambda}\left| {\boldsymbol \psi}_m  -\tilde{\boldsymbol \psi}_k^{\rm Pin}\right|
 }$ are due to signals' propagation from the antennas to the users.

Therefore,  ${\rm U}_m$'s signal-to-interference-plus-noise ratio (SINR) can be expressed as follows: 
     \begin{align}
 {\rm SINR}_m =\frac{ P\left|\sum^{K}_{k=1}h_{m,k} p_{m,k}\right|^2}{  P\sum_{i\neq m}\left|\sum^{K}_{k=1}h_{m,k} p_{i,k} \right|^2   + \sigma^2}. 
  \end{align}

The similarity between the considered pinching-antenna system and the conventional MISO interference channel can be illustrated by first  defining    $\mathbf{p}_m=\begin{bmatrix}
 p_{m,1}^* &\cdots &p_{m,K}^*
 \end{bmatrix}^T$,  $\mathbf{P}=\begin{bmatrix}
\mathbf{p}_1&\cdots &\mathbf{p}_M
 \end{bmatrix}^T$, and $\mathbf{h}_m=\begin{bmatrix}h_{m,1} &\cdots &h_{m,K} \end{bmatrix}^T$, and expressing $  {\rm SINR}_m$ as follows:
      \begin{align}\label{interference channels}
  {\rm SINR}_m =&\frac{P \left| \mathbf{h}_m^H\mathbf{p}_m\right|^2}{  P\sum_{i\neq m}\left|\mathbf{h}_m^H\mathbf{p}_i \right|^2   + \sigma^2}. 
  \end{align} 
  In order to clearly illustrate the key features of pinching-antenna systems with multiple waveguides, the special case of $M=K=2$ is focused on in the following subsections. 
  \vspace{-1em}  
\subsection{Existing Results for Two-User Interference Channels}  \label{subsection ula}
Two-user interference channels have been extensively studied in the literature, where the aim is to maximize the following two SINRs \cite{Cover1991,4558045,6553202}: 
       \begin{align}\label{sinrscover}
  {\rm SINR}_1 =\frac{\rho |\mathbf{h}_1^H\mathbf{p}_1|^2}{ \rho  |\mathbf{h}_1^H\mathbf{p}_2|^2 + 1}, \quad
    {\rm SINR}_2 =\frac{\rho|\mathbf{h}_2^H\mathbf{p}_2|^2}{   |\mathbf{h}_2^H\mathbf{p}_1|^2 + 1},
  \end{align} 
  where $\rho = \frac{P}{\sigma^2}$. The challenge of optimizing the SINRs of the two-user interference channels is that one user's SINR is improved at the price of another user's performance degradation. 
  
  \subsubsection{Practical Approaches} One low-complexity approach is termed MRC with $\mathbf{p}_m^{\rm MRC}=\frac{\mathbf{h}_m}{|\mathbf{h}_m^H\mathbf{h}_m|}$,  which means that the two users' SINRs are obtained as follows:
           \begin{align}
  {\rm SINR}_1^{\rm MRC} =\frac{\rho |\mathbf{h}_1 |^2}{ \rho \frac{|\mathbf{h}_1^H\mathbf{h}_2|^2}{|\mathbf{h}_2|^2} + 1}, 
  {\rm SINR}_1^{\rm MRC} =\frac{\rho|\mathbf{h}_2|^2}{ \rho \frac{|\mathbf{h}_2^H\mathbf{h}_1|^2}{|\mathbf{h}_1|^2} + 1}. 
  \end{align} 
  Another well-known approach is based on the ZF approach, where the beamforming vectors meet the following conditions:  $\mathbf{h}_1^H\mathbf{p}_2^{\rm ZF}=0$, $\mathbf{h}_2^H\mathbf{p}_1^{\rm ZF}=0$ and $|\mathbf{p}_m^{\rm ZF}|^2=1$, $m\in\{1,2\}$. By using the ZF approach, the two users' SINRs are obtained as follows:
        \begin{align}
  {\rm SINR}_1^{\rm ZF} =&\rho |\mathbf{h}_1^H\mathbf{p}_1^{\rm ZF}|^2, \quad 
    {\rm SINR}_2^{\rm ZF} =\rho |\mathbf{h}_2^H\mathbf{p}_2^{\rm ZF}|^2. 
  \end{align} 
MRC and ZF have their advantages and disadvantages. MRC can maximize the intended user's signal strength at the price of uncontrolled interference. ZF can completely suppress the co-channel interference but cannot maximize the intended user's signal strength as MRC can.  
 
 \begin{figure*}[t]\vspace{-2em}
{\small \begin{align}\label{pmk choice}
\begin{bmatrix} p_{1,1} & p_{2,1} \\p_{1,2} &p_{2,2} \end{bmatrix}= 
 \begin{bmatrix} 
\eta_1 \left| {\boldsymbol \psi}_1  -\tilde{\boldsymbol \psi}_1^{\rm Pin}\right|^{-1} e^{2\pi j \left(\frac{  1}{\lambda}\left| {\boldsymbol \psi}_1  -\tilde{\boldsymbol \psi}_1^{\rm Pin}\right|
  +\frac{1}{\lambda_g}\left| {\boldsymbol \psi}_0^{\rm Pin}  -\tilde{\boldsymbol \psi}_1^{\rm Pin}\right|
  \right)} &
 \eta_2  \left| {\boldsymbol \psi}_2  -\tilde{\boldsymbol \psi}_1^{\rm Pin}\right|^{-1} e^{2\pi j \left(\frac{  1}{\lambda}\left| {\boldsymbol \psi}_2  -\tilde{\boldsymbol \psi}_1^{\rm Pin}\right|
  +\frac{1}{\lambda_g}\left| {\boldsymbol \psi}_0^{\rm Pin}  -\tilde{\boldsymbol \psi}_1^{\rm Pin}\right|
  \right)} 
 \\ \eta_1 \left| {\boldsymbol \psi}_1  -\tilde{\boldsymbol \psi}_2^{\rm Pin}\right|^{-1} e^{2\pi j \left(\frac{  1}{\lambda}\left| {\boldsymbol \psi}_1  -\tilde{\boldsymbol \psi}_2^{\rm Pin}\right|
  +\frac{1}{\lambda_g}\left| {\boldsymbol \psi}_0^{\rm Pin}  -\tilde{\boldsymbol \psi}_2^{\rm Pin}\right|
  \right)}
 &\eta_2\left| {\boldsymbol \psi}_2  -\tilde{\boldsymbol \psi}_2^{\rm Pin}\right|^{-1} e^{2\pi j \left(\frac{  1}{\lambda}\left| {\boldsymbol \psi}_2  -\tilde{\boldsymbol \psi}_2^{\rm Pin}\right|
  +\frac{1}{\lambda_g}\left| {\boldsymbol \psi}_2^{\rm Pin}  -\tilde{\boldsymbol \psi}_m^{\rm Pin}\right|
  \right)} \end{bmatrix} .
\end{align}}\vspace{-2em}
\end{figure*}

 \subsubsection{An Upper Bound} A straightforward upper bound on the two users' SINRs can be obtained by considering the case, in which ${\rm U}_m$ can solely occupy the whole bandwidth. Therefore, the two users' SINRs can be upper bounded as follows\footnote{Alternatively, the upper bound in \eqref{upper bound} can be obtained by applying the Cauchy–Schwarz inequality to the numerator of the SINRs in \eqref{sinrscover}. }: 
           \begin{align}\label{upper bound}
  {\rm SINR}_1  \leq  \rho |\mathbf{h}_1 |^2 , \quad 
  {\rm SINR}_2  \leq  \rho |\mathbf{h}_2 |^2 ,
  \end{align} 
 which is generally not achievable by optimizing the beamforming vectors $\mathbf{p}_m$ only. 
\vspace{-1em}  
\subsection{Approaching the Upper Bound} \label{subsection achiving}
Two necessary conditions for realizing the upper bound are as follows:
\begin{itemize}
\item Phase-Matching Condition: For ${\rm U}_m$, $1\leq m\leq M$,  the phase of $p_{m,m}$ matches the phase of $h_{m,m}$, i.e., the difference between the phases of $p_{m,m}$  and  $h_{m,m}$ must be multiples of $2\pi$ and hence the numerator of ${\rm SINR}_m $ can be $ |\mathbf{h}_m |^2$, $m\in \{1,2\}$.

\item Orthogonality Condition: Each user does not experience interference from the other user, which ensures the denominator of ${\rm SINR}_m $ is $1$, $m\in \{1,2\}$,  i.e., 
\begin{align}
 h_{1,1}p_{2,1}+h_{1,2}p_{2,2} =0,   h_{2,1}p_{1,1}+h_{2,2}p_{1,2}=0. 
\end{align}

\end{itemize} 
   
In the pinching-antenna system, by adjusting the antenna locations, the channels $h_{m,k}$ also become configurable parameters, which means that the upper bound in \eqref{upper bound} might be achievable. We note that the upper bound in \eqref{upper bound} is a function of the antenna locations, i.e., the upper bound changes when the antennas are moved. However, if the two conditions are met by moving the antennas in a micro-meter length scale, e.g., by a few wavelengths in millimeter or terahertz networks, the upper bound does not change significantly. In other words, the two conditions are the necessary and sufficient conditions if they can be met by moving the antennas on a micro-meter scale. 


\subsubsection{Feasibility Analysis}
There are two sets of parameters to be designed, namely $p_{m,k}$ and $\tilde{\boldsymbol \psi}_m^{\rm Pin}$. To facilitate the feasibility analysis, the beamforming coefficients will be designed first by assuming that the antenna locations are fixed. 

Recall that the phase of $h_{m,m}$ is determined by the term,  $e^{-2\pi j \left(\frac{  1}{\lambda}\left| {\boldsymbol \psi}_m  -\tilde{\boldsymbol \psi}_k^{\rm Pin}\right|
  +\frac{1}{\lambda_g}\left| {\boldsymbol \psi}_0^{\rm Pin}  -\tilde{\boldsymbol \psi}_m^{\rm Pin}\right|
  \right)}$.  Therefore, for fixed $\tilde{\boldsymbol \psi}_m^{\rm Pin}$,   $p_{m,n}$ can be chosen as shown in \eqref{pmk choice} at the top of this page,
where $\eta_m$ is the power normalization parameter. In particular, the constraint that  $|p_{m,1}|^2+|p_{m,2}|^2=1$ leads to $
\eta_1=\left(\left| {\boldsymbol \psi}_1  -\tilde{\boldsymbol \psi}_1^{\rm Pin}\right|^{-2}   +  \left| {\boldsymbol \psi}_1  -\tilde{\boldsymbol \psi}_2^{\rm Pin}\right|^{-2}\right)^{-\frac{1}{2}}  $ and  $
\eta_2=\left( \left| {\boldsymbol \psi}_2  -\tilde{\boldsymbol \psi}_1^{\rm Pin}\right| ^{-2}  + \left| {\boldsymbol \psi}_2  -\tilde{\boldsymbol \psi}_2^{\rm Pin}\right| ^{-2}\right)^{-\frac{1}{2}}  $.  
 
The choices of $p_{m,n}$   in \eqref{pmk choice} have two benefits. One is that the phase-matching condition is satisfied. The other is that the orthogonality condition can be simplified. In particular, $ h_{1,1}p_{2,1}+h_{1,2}p_{2,2} =0$ simplifies as follows:
\begin{align}\label{dsdf}
\frac{\sqrt{\eta} e^{-j\frac{2\pi }{\lambda}\left| {\boldsymbol \psi}_1  -\tilde{\boldsymbol \psi}_1^{\rm Pin}\right|}}{  \left| {\boldsymbol \psi} _1 -\tilde{\boldsymbol \psi}_1^{\rm Pin}\right|}  \eta_2  \frac{ e^{j\frac{2\pi }{\lambda}\left| {\boldsymbol \psi}_2  -\tilde{\boldsymbol \psi}_1^{\rm Pin}\right|} }{\left| {\boldsymbol \psi}_2  -\tilde{\boldsymbol \psi}_1^{\rm Pin}\right|}\\\nonumber
+\frac{\sqrt{\eta} e^{-j\frac{2\pi }{\lambda}\left| {\boldsymbol \psi}_1  -\tilde{\boldsymbol \psi}_2^{\rm Pin}\right|}}{  \left| {\boldsymbol \psi} _1 -\tilde{\boldsymbol \psi}_2^{\rm Pin}\right|} \eta_2 \frac{e^{j\frac{2\pi }{\lambda}\left| {\boldsymbol \psi}_2  -\tilde{\boldsymbol \psi}_2^{\rm Pin}\right|}}{\left| {\boldsymbol \psi}_2  -\tilde{\boldsymbol \psi}_2^{\rm Pin}\right|} &=0,
\end{align}
where it is interesting to note that the phase shifts caused by the signals passing through the waveguide are eliminated.   
This equality in \eqref{dsdf} can be further written as follows:
 \begin{align}\label{dsdf33}
&  e^{-j\frac{2\pi }{\lambda}\left( \left| {\boldsymbol \psi}_1  -\tilde{\boldsymbol \psi}_1^{\rm Pin}\right|-\left| {\boldsymbol \psi}_2-  \tilde{\boldsymbol \psi}_1^{\rm Pin}\right|-\left| {\boldsymbol \psi}_1  -\tilde{\boldsymbol \psi}_2^{\rm Pin}\right|+\left| {\boldsymbol \psi}_2  -\tilde{\boldsymbol \psi}_2^{\rm Pin}\right|
 \right)}  =
\\\nonumber &-  {\left| {\boldsymbol \psi}_1  -\tilde{\boldsymbol \psi}_1^{\rm Pin}\right|\left| {\boldsymbol \psi}_2  -\tilde{\boldsymbol \psi}_1^{\rm Pin}\right| } \left| {\boldsymbol \psi}_1  -\tilde{\boldsymbol \psi}_2^{\rm Pin}\right|^{-1}\left| {\boldsymbol \psi}_2  -\tilde{\boldsymbol \psi}_2^{\rm Pin}\right|^{-1}  \hspace{-0.5em},
\end{align}
which leads to the following two constraints:
 \begin{align} \nonumber
\text{Constraint 1:}& \left| {\boldsymbol \psi}_1  -\tilde{\boldsymbol \psi}_1^{\rm Pin}\right|-\left| {\boldsymbol \psi}_2-  \tilde{\boldsymbol \psi}_1^{\rm Pin}\right|\\  &-\left| {\boldsymbol \psi}_1  -\tilde{\boldsymbol \psi}_2^{\rm Pin}\right|+\left| {\boldsymbol \psi}_2  -\tilde{\boldsymbol \psi}_2^{\rm Pin}\right|
=\frac{k\lambda}{2},\label{dsdf33x1}
\end{align}
and 
\begin{align}\label{dsdf33x2}
\text{Constraint 2:}\quad  \frac{\left| {\boldsymbol \psi}_1  -\tilde{\boldsymbol \psi}_1^{\rm Pin}\right|\left| {\boldsymbol \psi}_2  -\tilde{\boldsymbol \psi}_1^{\rm Pin}\right| }{\left| {\boldsymbol \psi}_1  -\tilde{\boldsymbol \psi}_2^{\rm Pin}\right|\left| {\boldsymbol \psi}_2  -\tilde{\boldsymbol \psi}_2^{\rm Pin}\right|} =1 ,
\end{align}
where $k$ must be an odd integer. 
It can be verified that  the constraint $ h_{2,1}p_{1,1}+h_{2,2}p_{1,2}=0$ leads to the same constraints as shown in \eqref{dsdf33x1} and \eqref{dsdf33x2}.  

Depending on the deployment of the users and the waveguides, the two constraints in \eqref{dsdf33x1} and \eqref{dsdf33x2} can be met by moving the antennas in a micro-meter scale, as to be shown in Section \ref{section5}. In other words, the use of pinching antennas can make the upper bound achievable.  We have yet to obtain a rigorous analysis for the impact of the user/waveguide deployment on the feasibility of the two constraints, which is an important direction for future research. However, the special case provided in the following section gives some insight into the feasibility of the two constraints.

\subsubsection{A Special Case Which Guarantees Constraints in \eqref{dsdf33x1} and \eqref{dsdf33x2}} There exist communication scenarios where it is always feasible to find antenna locations satisfying the constraints in \eqref{dsdf33x1} and \eqref{dsdf33x2}. For example, consider that the two users are located on the x-axis, i.e., their coordinates are $(x_1,0,0)$ and $(x_2,0,0)$. The two waveguides are placed as shown in Fig. \ref{fig3}.  Without loss of generality, assume that $x_1<x_2$. For an arbitrary $\Delta\geq 0$,  if the two pinching antennas are placed at $\left(x_1+\Delta, \frac{D}{3}, d\right)$ and $\left(x_2-\Delta, -\frac{D}{3}, d\right)$, it is straightforward to verify that the following holds:
\begin{align}
\left| {\boldsymbol \psi}_1  -\tilde{\boldsymbol \psi}_1^{\rm Pin}\right|=\left| {\boldsymbol \psi}_2  -\tilde{\boldsymbol \psi}_2^{\rm Pin}\right|,
\left| {\boldsymbol \psi}_2  -\tilde{\boldsymbol \psi}_1^{\rm Pin}\right|=\left| {\boldsymbol \psi}_1  -\tilde{\boldsymbol \psi}_2^{\rm Pin}\right|,
\end{align}
 which guarantees the feasibility of constraint \eqref{dsdf33x2}.  The above equality also simplifies  constraint \eqref{dsdf33x1} as follows:
 \begin{align}  \label{kfeasible}
\frac{k\lambda}{4}= \left| {\boldsymbol \psi}_1  -\tilde{\boldsymbol \psi}_1^{\rm Pin}\right|-\left| {\boldsymbol \psi}_2-  \tilde{\boldsymbol \psi}_1^{\rm Pin}\right| \triangleq f(\tilde{x}_1^{\rm Pin}).
\end{align}
We note that $f(x)$ can be further expressed as follows:
 \begin{align}  \nonumber
f(x) = \sqrt{(x-x_1)^2+\frac{D^2}{9}+d^2} - \sqrt{(x-x_2)^2+\frac{D^2}{9}+d^2},
\end{align}
whose first-order derivative is given by
  \begin{align}  
f'(x) = 2(x-x_1) a_1 - 2(x-x_2) a_2\geq 0,
\end{align}
for $x_1\leq x\leq \frac{x_2-x_1}{2}$, 
where $a_1=\left((x-x_1)^2+\frac{D^2}{9}+d^2\right)^{-\frac{1}{2}}$ and $a_2=\left((x-x_2)^2+\frac{D^2}{9}+d^2\right)^{-\frac{1}{2}}$. Therefore, for $x_1\leq \tilde{x}_1^{\rm Pin}\leq \frac{x_2-x_1}{2}$, $f(\tilde{x}_1^{\rm Pin})$ is a monotonically increasing function with the following range:
{\small \begin{align}  \label{range}
 &\sqrt{ \frac{D^2}{9}+d^2} - \sqrt{(x_1-x_2)^2+\frac{D^2}{9}+d^2}\leq  f(\tilde{x}_1^{\rm Pin})\leq 0. 
\end{align}}
In other words, any value in the range shown in \eqref{range} is achievable by adjusting $\tilde{x}_1^{\rm Pin}$. 
Therefore, for large $|x_1-x_2|$, the range in \eqref{range} must be much larger than $\lambda$, which means that a feasible choice of $k$ to satisfy \eqref{kfeasible} can be found. As shown in Section \ref{section5}, multiple feasible choices of $k$ exist, even if the two users are not located on a line parallel to the waveguides. 
 
\subsubsection{A Search-Based Algorithm to Approach the Upper Bound}
While the steps in the previous section can be used for the feasibility analysis, they cannot be directly used to find desirable $\tilde{\boldsymbol \psi}_n^{\rm Pin}$ and $p_{m,k}$. 
In order to demonstrate the capability of pinching antennas to achieve the upper bound, a search is conducted to find the ideal locations of the pinching antennas, as shown in Algorithm \ref{algorithm}. In brief, the search-based algorithm enumerates all potential locations of the pinching antennas. Recall from the conducted feasibility study that if the ideal antenna locations are found, low-complexity approaches, e.g., ZF and MRC, should achieve the upper bound. Therefore, in each iteration of the search, the ZF approach is used to obtain the corresponding beamforming vectors, as well as the users' SINRs. Various metrics can be used for the location selection, and our simulation results show that the use of the max-min criterion shown in Algorithm \ref{algorithm} is sufficient for the achievability of the upper bound.

 \begin{algorithm}[t]
\caption{A Search-Based Algorithm}

 \begin{algorithmic}[1]
 
\State Build two vectors $\mathbf{v}_1$ and $\mathbf{v}_2$ collecting the locations to be searched; denote the length of $\mathbf{v}_m$ by $N_{v_m}$ 
\For{ $n_1=1 : N_{v_1}$ }
\For{ $n_2=1 : N_{v_2}$ }
\State $\tilde{x}_1^{\rm Pin}=\mathbf{v}_1[n_1]$ and $\tilde{x}_2^{\rm Pin}=\mathbf{v}_2[n_2]$. 
\State  Generate the new coordinates of the  antennas 

 \quad - $\tilde{\boldsymbol \psi}_1^{\rm Pin}=\left(\tilde{x}_1^{\rm pin}, \frac{D}{3}, d\right)$ 
 
 \quad - $\tilde{\boldsymbol \psi}_2^{\rm Pin}=\left(\tilde{x}_2^{\rm pin}, -\frac{D}{3}, d\right)$

\State Use $\tilde{\boldsymbol \psi}_m^{\rm Pin}$ to generate $\mathbf{h}_m$ and $\mathbf{H}=\begin{bmatrix} \mathbf{h}_1& \mathbf{h}_2 \end{bmatrix}$ 

\State  Obtain  $\mathbf{p}_m$ as   the ZF vectors of $\mathbf{H}$    
 
 \State Find ${\rm SINR}_m(n_1,n_2)$, $m\in\{1,2\}$, in \eqref{sinrscover}
     
  \EndFor
 
 \EndFor
 \State \quad \quad \textbf{end}
 
\State \textbf{end}

\State Obtain $(n_1^*,n_2^*)=\arg\max \min \{{\rm SINR}_m(n_1,n_2), m\in\{1,2\}\}$.

\State Output $\mathbf{p}_m^*$ and  $\tilde{\boldsymbol \psi}_m^{\rm Pin *}$ by using $(n_1^*,n_2^*)$
 \end{algorithmic}\label{algorithm}
\end{algorithm}  

%

\vspace{-1em}

\section{Simulation Results}\label{section5}
In this section, computer simulations are used to evaluate the performance of the pinching-antenna system. For illustration purposes, the noise power is set as $-90$ dBm,  $d=3$ m, $f_c=28$ GHz, $\lambda_{\rm cut}=10$ GHz, $\tilde{\Delta}=\frac{\lambda}{2}$ and $n_{\rm eff}=1.4$ \cite{microwave}. The three scenarios considered in Sections  \ref{section2}, \ref{section3}, and \ref{section4} are studied in the following three subsections, respectively.   \vspace{-1em}  
\subsection{ The Single-Pinching-Antenna Single-Waveguide Case}\label{-1em}
\begin{figure}[t] \vspace{-1em}
\begin{center}
\subfigure[Case I]{\label{fig4}\includegraphics[width=0.35\textwidth]{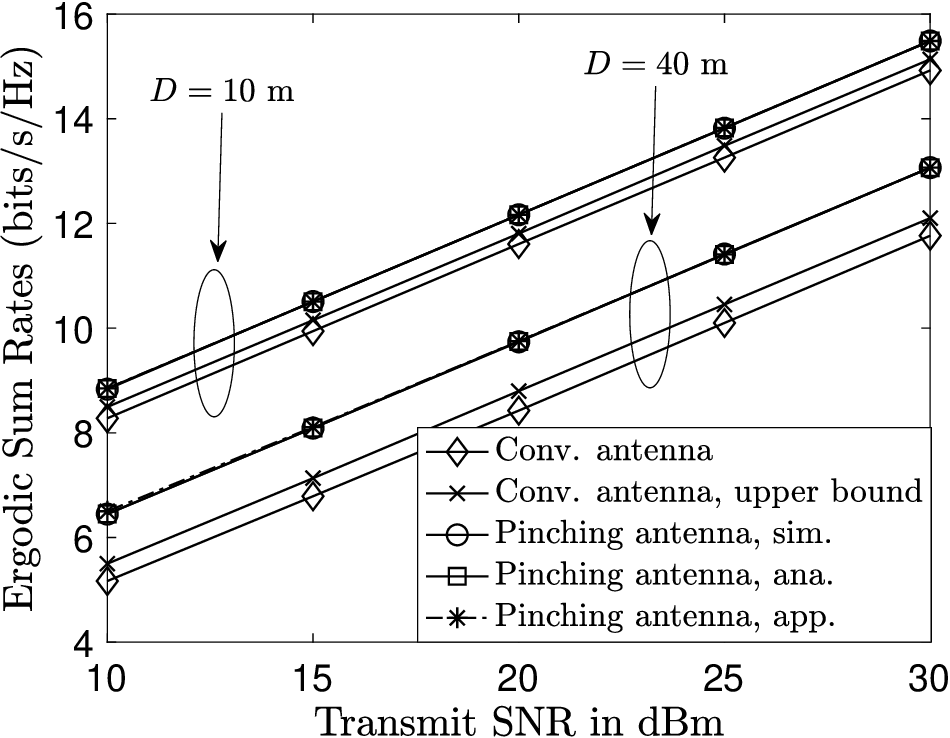}} 
\subfigure[Case II]{\label{fig5}\includegraphics[width=0.35\textwidth]{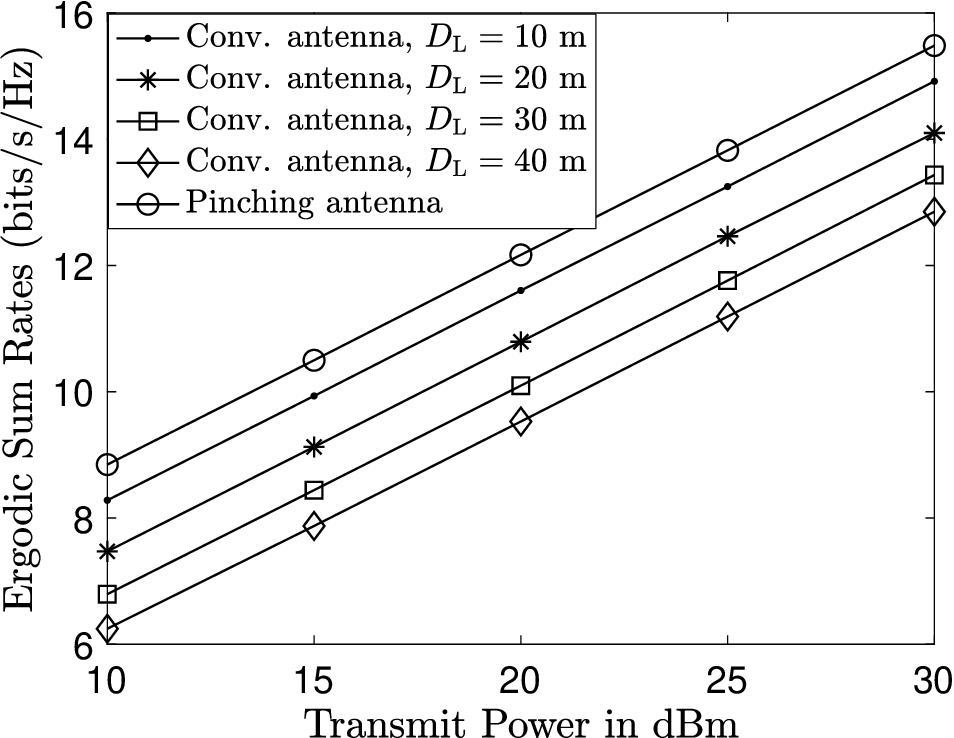}} \vspace{-1em}
\end{center}
\caption{Ergodic sum rates achieved by the considered schemes, with a single pinching antenna and a single waveguide. The coordinates of the points on the waveguide are $(\tilde{x}^{\rm pin}, 0, d)$. In Case I, the users are uniformly distributed within a square, with side length $D$ and its center at $(0,0,0)$. The analytical results are based on \eqref{sum rate pinx3}, and the approximation results are based on \eqref{approxm pinching an43}. The upper bound of the conventional antenna is based on \eqref{conv approximation}.  In Case II, the users are uniformly distributed within a rectangle, with its two side lengths being $D$ and $D_{\rm L}$, where $D=10$ m.    \vspace{-1em} }\label{fig45}\vspace{-1.2em}
\end{figure}

In Fig. \ref{fig4}, the users are assumed to be uniformly distributed in a square, with side length $D$ and its center at $(0,0,0)$. As shown in Fig. \ref{fig4}, the use of a pinching antenna can achieve an ergodic sum rate larger than that of the conventional antenna system. This performance gain is due to the fact that the use of pinching antennas can reduce the users' path losses. Fig. \ref{fig4} also shows that the performance gain of the pinching antenna over the conventional one is enlarged by increasing the size of the area, which demonstrates the unique capability of the pinching-antenna system to create strong LoS links and mitigate large-scale path loss. Furthermore, the simulation results shown in Fig. \ref{fig4} also verify the accuracy of the analytical results developed in Section \ref{section2}. 

Unlike Fig. \ref{fig4}, Fig. \ref{fig5} assumes that the users are randomly distributed in a rectangular-shaped area, where the waveguide is placed parallel to the long side of the rectangle. Fig. \ref{fig5} shows that the performance gain of the pinching antenna over the conventional one can be increased significantly if the length of the long side of the rectangle is increased. The reason is that by increasing the long side of the rectangle, a user's distance to the center of the rectangle is increased, which means that with the conventional antenna, the user experiences a larger path loss on average. However, the pinching antenna can be flexibly placed next to the user, e.g., $\boldsymbol{\psi}^{\rm pin}_m$ shown in Fig. \ref{fig1}, which means that the users' path losses remain the same, as long as the width of the rectangle is fixed. 
\vspace{-1em}  
\subsection{The  Multiple-Pinching-Antenna Single-Waveguide Case}\label{-1em}
\begin{figure}[t]\centering \vspace{-2em}
    \epsfig{file=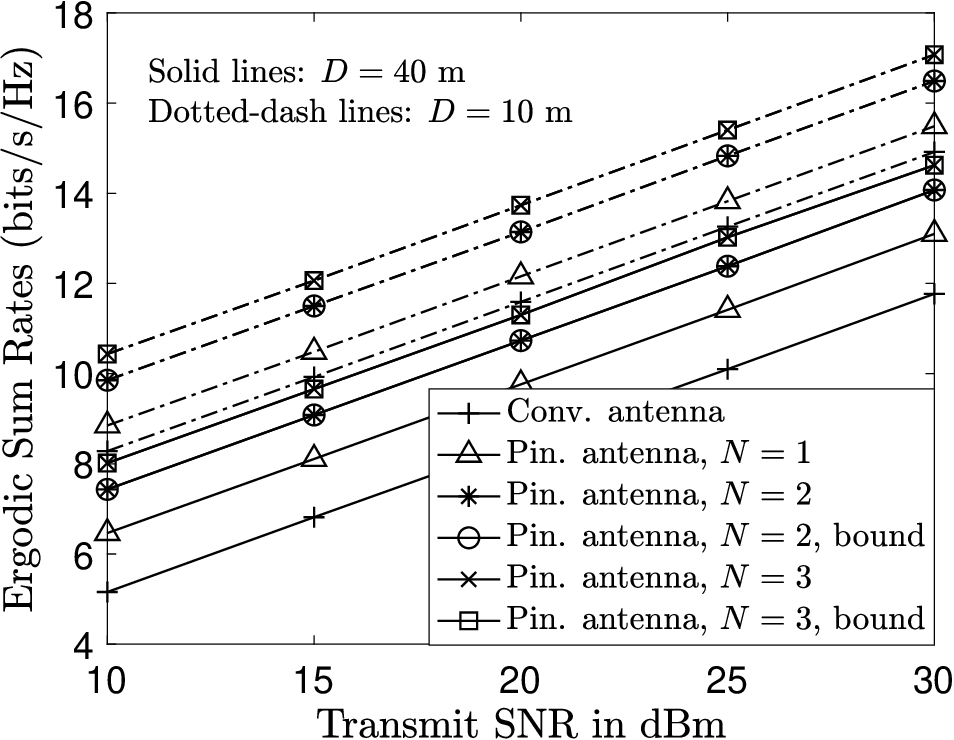, width=0.35\textwidth, clip=}\vspace{-0.5em}
\caption{Ergodic sum rates achieved by the considered schemes, with $N$ pinching antennas and a single waveguide. The users are uniformly distributed within a square, with side length $D$ and its center at $(0,0,0)$, and the coordinates of the points on the waveguide are $(\tilde{x}^{\rm pin}, 0, d)$. The upper bound curves are based on the result in \eqref{mulpa1wg}. The locations of the $N$ pinching antennas are obtained by the search algorithm proposed in Section \ref{subsection NP}. 
  \vspace{-1em}    }\label{fig6}   \vspace{-1em} 
\end{figure}

Fig. \ref{fig6} focuses on the scenario in which TDMA is used to serve the users. As can be seen from the figure, by increasing the number of pinching antennas, the performance gain of the pinching-antenna system over the conventional one can be increased significantly. We note that, unlike conventional antennas, pinching antennas can be flexibly deployed, and increasing the number of pinching antennas incurs almost no additional cost \cite{pinching_antenna2}. The figure also demonstrates that the upper bound shown in \eqref{mulpa1wg} can be achieved by the search algorithm developed at the end of Section \ref{subsection NP}.

     \begin{figure}[t]\centering \vspace{-0em}
    \epsfig{file=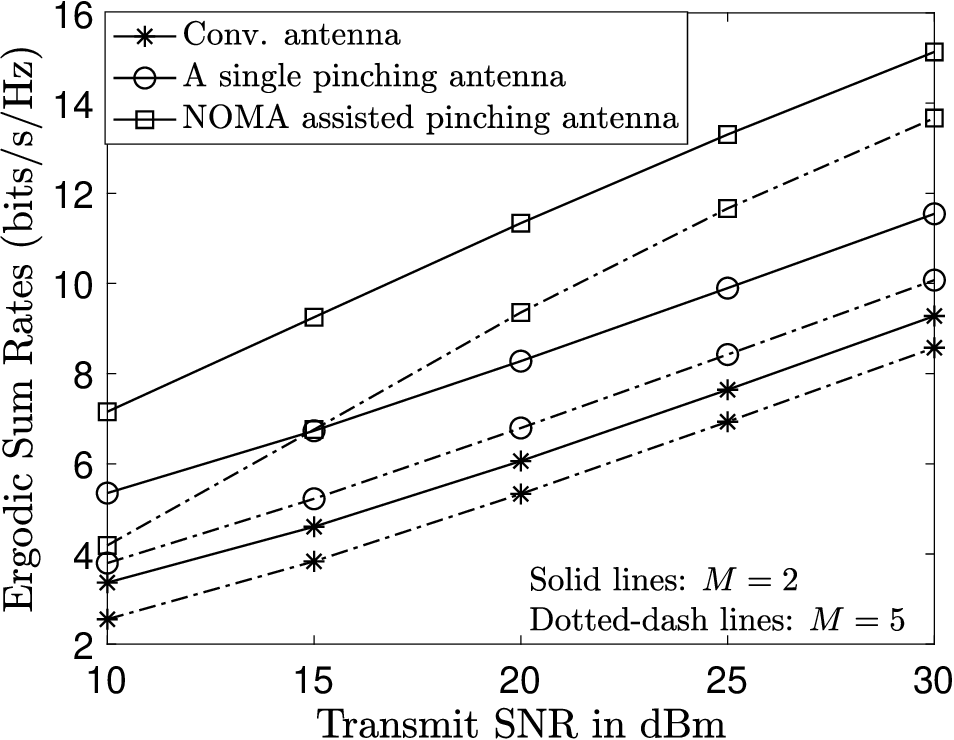, width=0.35\textwidth, clip=}\vspace{-0.5em}
\caption{Ergodic sum rates achieved by pinching-antenna assisted NOMA, with $N=M$ pinching antennas and a single waveguide.  The coordinates of the points on the waveguide are $(x_1^{\rm pin}, 0, d)$.  Each user is uniformly distributed in a square with side length $D=2$ m, which is denoted by $A_{m}$, as shown in Fig. \ref{fig2}. The coordinates of the center of $A_M$ are $(-10,0,0)$. The coordinates of the centers of $A_m$, $1\leq m \leq M-1$, are $((M-m)D_m, (M-m)D_m, 0)$, where $D_m=20$ m for $1\leq m < M$.  The NOMA power coefficients are obtained by first building a vector $\mathbf{b}=\begin{bmatrix}2M+1&\cdots&3&1 \end{bmatrix}$ and $\alpha_m=\frac{\mathbf{b}[m]}{
\mathbf{b}^T\mathbf{1}_M}$, where $\mathbf{1}_m$ is an $m\times 1$ all-one vector.
  \vspace{-1em}    }\label{fig7}   \vspace{-1.2em} 
\end{figure}

\begin{figure}[t]\centering \vspace{-2em}
    \epsfig{file=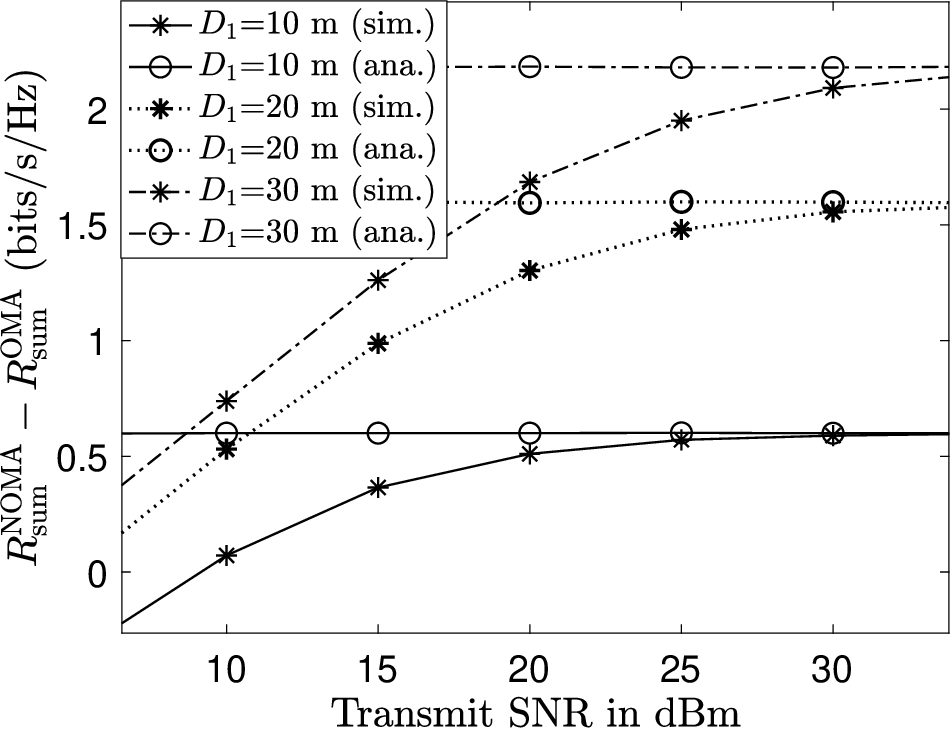, width=0.35\textwidth, clip=}\vspace{-0.5em}
\caption{Performance gain of NOMA over OMA,  with $N$ pinching antenna and a single waveguide.     $M=N=2$, and $D_2=10$ m. The analytical results are based on \eqref{gapgain}.  The other parameters are the same as for Fig. \ref{fig7}.   
  \vspace{-1em}    }\label{fig8ddd}   \vspace{-0.1em} 
\end{figure}

With multiple pinching antennas activated on a single waveguide, multiple users can be simultaneously served by applying NOMA, as shown in Section \ref{subsection NOMA}. In Fig. \ref{fig7}, the sum rate is used as the metric to evaluate the performance achieved by the proposed NOMA-assisted pinching-antenna system. As can be seen from the figure, the performance gain of the pinching-antenna system over the conventional one can be significantly increased by applying the NOMA principle. We note that the sum rate of the $M=5$ case is smaller than that of $M=2$. This decrease is due to the considered user deployment strategy, i.e., additional users are deployed in areas far away from the waveguide and hence suffer severe path losses.  Compared to Fig. \ref{fig4}, the performance gap between the single-pinching-antenna case and the conventional one shown in Fig. \ref{fig7} is larger, which is due to the fact that the weak and strong users are randomly deployed in two different areas as shown in Fig. \ref{fig2}. In Fig. \ref{fig8ddd}, the NOMA-assisted pinching-antenna system proposed in Section \ref{subsection NOMA} is compared to the OMA-assisted pinching-antenna system proposed in Section \ref{subsection NP}. As can be seen from the figure, the NOMA system can outperform the OMA one, and the performance gain of NOMA over OMA, i.e., $R_{\rm sum}^{\rm NOMA}- R_{\rm sum}^{\rm OMA}$, is increased if the two users' channel conditions become more different. Fig. \ref{fig8ddd} also shows the accuracy of the approximation reported in \eqref{gapgain} at high SNR. We note that the approximation in \eqref{gapgain} requires $\frac{P  }{  \left| {\boldsymbol \psi} _1 -{\boldsymbol \psi}_1^{\rm Pin}\right|^2N}   $ to be large, which is the reason why in Fig. \ref{fig8ddd} the accuracy is better with smaller $D_1$.   

     \begin{figure}[t]\centering \vspace{-0em}
    \epsfig{file=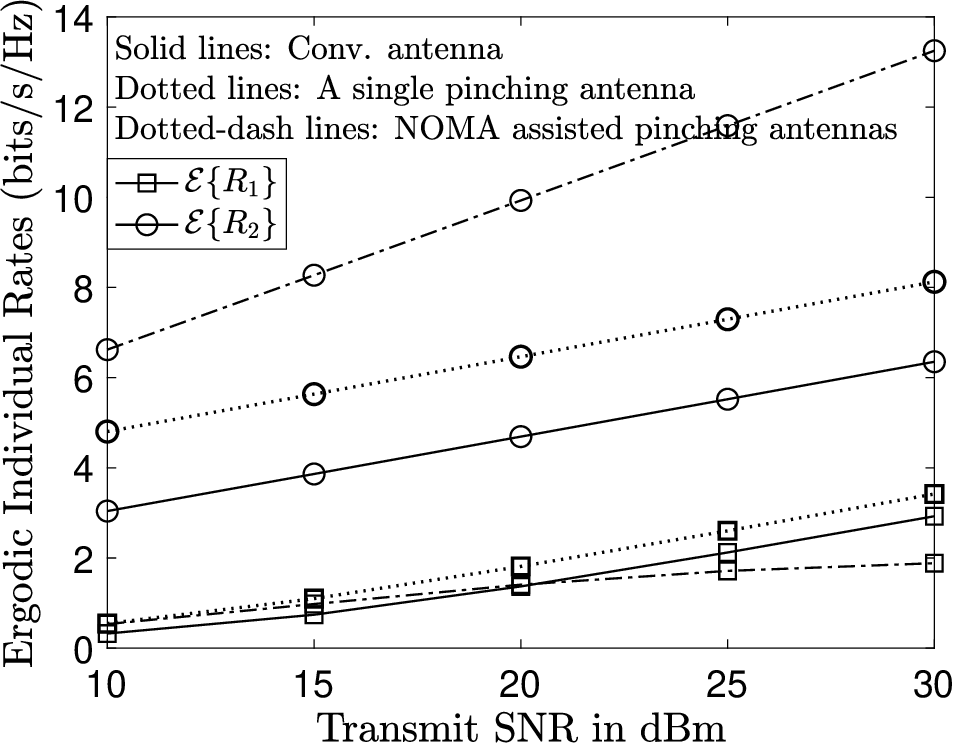, width=0.35\textwidth, clip=}\vspace{-0.5em}
\caption{The users' individual data rates achieved by pinching-antenna assisted NOMA, with $N$ pinching antennas and a single waveguide.     $M=N=2$. The other parameters are based on the same choices as Fig. \ref{fig7}.   
  \vspace{-1em}    }\label{fig8}   \vspace{-0.5em} 
\end{figure}

Figs. \ref{fig7} and \ref{fig8ddd} show that the use of NOMA can increase the sum rate compared to the benchmarking schemes. However, we note that this sum rate increase is at the price of the weak users' data rates. To clearly illustrate this effect, the users' individual data rates are investigated in Fig. \ref{fig8}, where a two-user NOMA scenario is focused on. As can be seen from the figure, by applying the NOMA principle, the strong user benefits the most since its data rate is increased significantly compared to the single-pinching-antenna case. However, the weak user suffers a reduction of the data rate, particularly at high SNR. This is due to the fact that the weak user treats its partner's signal as noise, which means that at high SNR, its data rate shown in \eqref{um1 rate2} approaches a constant, i.e., $ R_1 \approx  \log_2  \left(1+ \frac{\frac{\eta  }{  \left| {\boldsymbol \psi} _1 -{\boldsymbol \psi}_1^{\rm Pin}\right|^2}  \frac{P}{N}\alpha_1}{ \frac{\eta  }{  \left| {\boldsymbol \psi} _1 -{\boldsymbol \psi}_1^{\rm Pin}\right|^2}   \frac{ P}{N} \alpha_2+\sigma^2} \right)\approx  \log_2  \left(1+ \frac{\alpha_1}{\alpha_2} \right)$.

 \vspace{-1em}  
\subsection{The Multi-Pinching-Antenna Multi-Waveguide Case}\label{-1em}

     \begin{figure}[t] \vspace{-2em}
\begin{center}
\subfigure[Pinching antennas vs fixed antennas]{\label{fig9a}\includegraphics[width=0.35\textwidth]{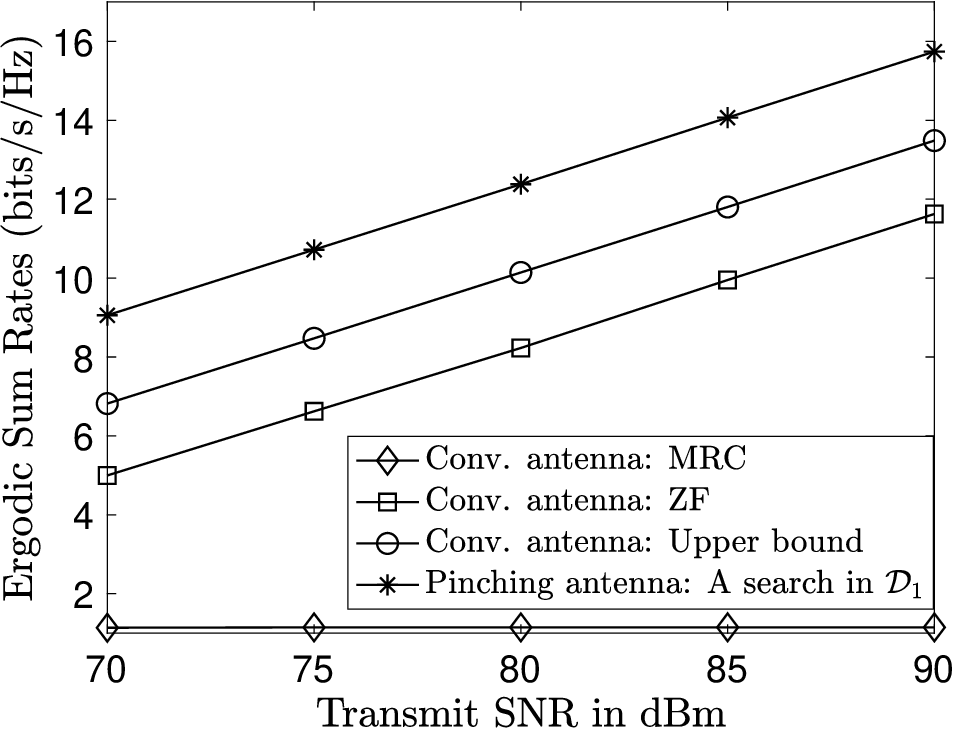}} 
\subfigure[ Low-complexity pinching-antenna methods]{\label{fig9b}\includegraphics[width=0.35\textwidth]{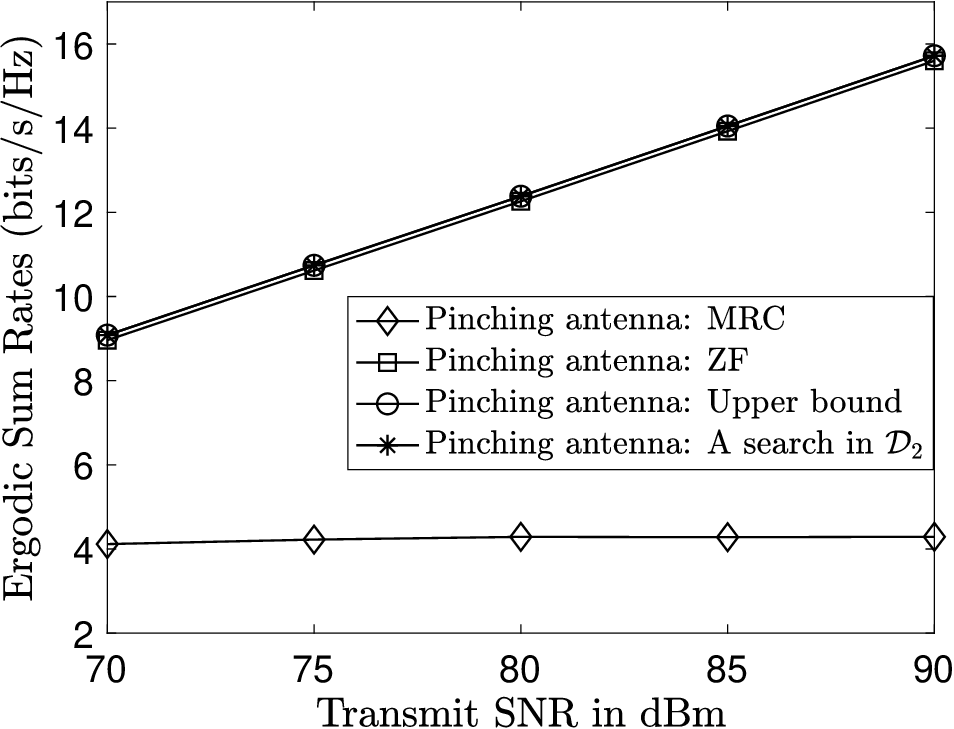}} \vspace{-1em}
\end{center}
\caption{Ergodic sum rates achieved by the considered transmission schemes, with two pinching antennas and two waveguides. Consider a square with side length $D=20$ m and its center at $(0,0,0)$. The coordinates of the points on the two waveguides are $(x_1^{\rm pin}, \frac{D}{3}, d)$ and $(x_2^{\rm pin}, -\frac{D}{3}, d)$, respectively, i.e., the waveguides divide the square into three rectangles. ${\rm U}_1$ is uniformly distributed within the upper rectangle, and  ${\rm U}_2$ is uniformly distributed within the lower rectangle. The two conventional antennas are placed at $\left( \frac{\lambda}{4},0,d\right)$ and $\left( -\frac{\lambda}{4},0,d\right)$, respectively.  $\mathcal{D}_1$ contains all the points on the two waveguides, and $\mathcal{D}_2=[x_1^{\rm pin}-10\lambda, x_1^{\rm pin}+10\lambda]\cup[x_2^{\rm pin}-10\lambda, x_2^{\rm pin}+10\lambda]  $, i.e.,  $\mathcal{D}_2$ contains the locations in close proximity to $\boldsymbol{\psi}_m^{\rm Pin}$. \vspace{-1em} }\label{fig9}\vspace{-1.1em}
\end{figure}

     \begin{figure}[t] \vspace{-2em}
\begin{center}
\subfigure[Case I]{\label{fig10a}\includegraphics[width=0.35\textwidth]{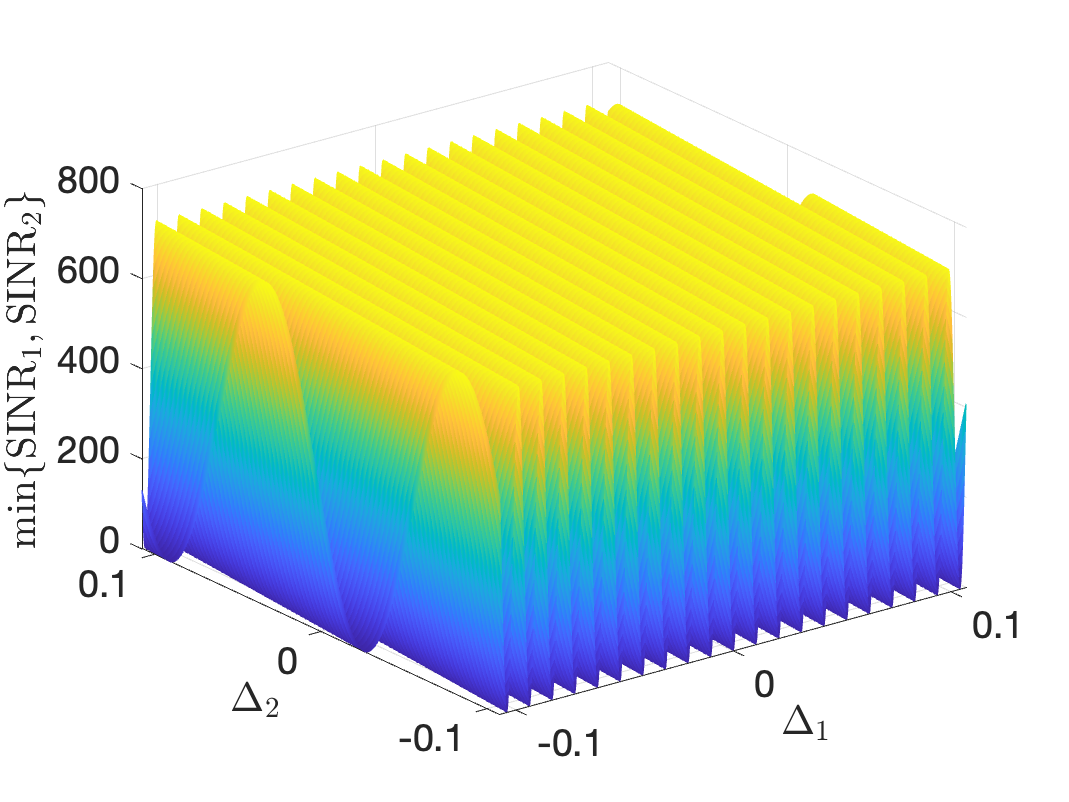}} 
\subfigure[Case II]{\label{fig10b}\includegraphics[width=0.35\textwidth]{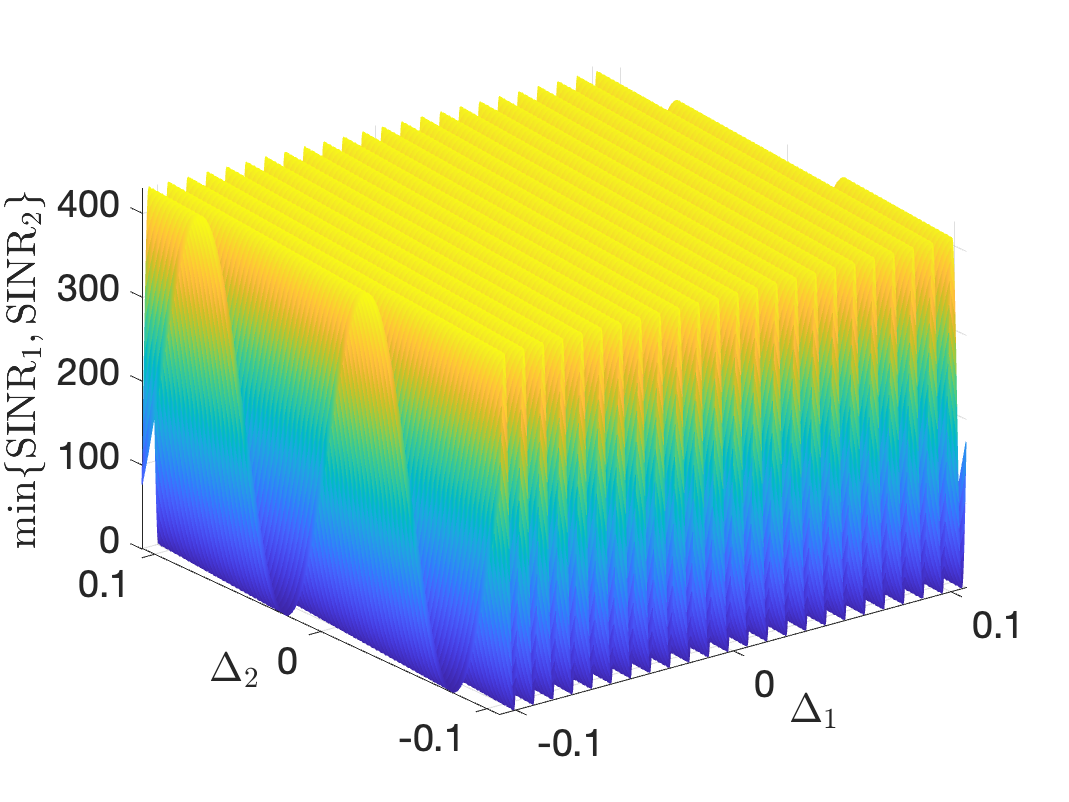}} \vspace{-1em}
\end{center}
\caption{Performance achieved by the low-complexity pinching-antenna schemes,  with two pinching antenna and two waveguides, where ${\rm SINR}_{\min}=\min\{{\rm SINR}_1, {\rm SINR}_2\}$. The users are uniformly distributed within a square, with side length $D=20$ m and its center at $(0,0,0)$, and the two shown cases are obtained for two random channel realizations. The $m$-th antenna is placed as follows: $\tilde{x}_m^{\rm Pin}= {x}_m^{\rm Pin}+\Delta_m$.  The other parameters are the same as for Fig. \ref{fig9}.   \vspace{-1em} }\label{fig10}\vspace{-1em}
\end{figure}

This subsection focuses on the case where two pinching antennas are activated on two waveguides and employed to serve two users, i.e., $N=K=M=2$. The users' locations are described in the caption of Fig. \ref{fig9}, which ensures that ${\rm U}_m$ is to be served by the $m$-th waveguide, and facilitates the discussion of the achievability of the upper bound. Fig. \ref{fig9a} shows the performance of the pinching-antenna system achieved by searching all the possible antenna locations (denoted by $\mathcal{D}_1$).  Three benchmarking schemes are used in Fig. \ref{fig9a}, including MRC, ZF, and the upper bound in \eqref{upper bound} based on the locations of the conventional antennas. As can be observed from Fig. \ref{fig9a}, the use of pinching antennas yields a significant performance gain over the benchmarking schemes. 

In Fig. \ref{fig9b}, a low-complexity search is conducted by focusing on locations in close proximity to $\boldsymbol{\psi}_m^{\rm Pin}$, $m\in\{1,2\}$ (denoted by $\mathcal{D}_2$). One observation from Fig. \ref{fig9b} is that the search yields the same performance as the upper bound corresponding to $\boldsymbol{\psi}_m^{\rm Pin}$. This observation is significant since it verifies that the upper bound can be achieved and the two constraints in \eqref{dsdf33x1} and \eqref{dsdf33x2} can be realized with micro-meter antenna movements. Another important observation from Fig. \ref{fig9} is that the two searches in the two subfigures achieve the same performance, which indicates that the optimal locations of the pinching antennas are very close to $\boldsymbol{\psi}_m^{\rm Pin}$. This observation also motivates the low-complexity approach to first place the pinching antennas next to their associated users, i.e., $\tilde{x}_m^{\rm pin} = x_m^{\rm pin}$,  $m\in \{1,2\}$, and then apply low-complexity beamforming methods, such as ZF and MRC. Fig. \ref{fig9b} shows that the performance gap between the conducted search and ZF is insignificant, which means that an exhaustive search can be avoided with a slight performance loss.  

We note that the achievability of the upper bound of the MISO interference channel depends on the user/waveguide deployment. For the case considered in Fig. \ref{fig9}, i.e., the users are uniformly distributed in two separated rectangles,  our simulation results indicate that the upper bound is always achievable. However, the upper bound achievability is not always guaranteed, as shown in Fig. \ref{fig10} and Table \ref{table1}, where the users are uniformly distributed within the same square, with side length $D$ and its center at $(0,0,0)$, and the two cases shown are obtained for two random channel realizations. As can be seen from Table \ref{table1}, the upper bound is achievable for the first case but not for the second one. Fig. \ref{fig10} confirms that there are multiple optimal pinching antenna locations that achieve the same performance. A rigorous study to identify the impact of the user/waveguide deployment on the achievability of the upper bound is an important direction for future research.

\section{Conclusions}\label{section6}
This paper has focused on a new type of flexible-antenna technology, termed pinching antennas. Analytical results were first developed for the case with a single pinching antenna and a single waveguide, where the capability of pinching-antenna systems to mitigate large-scale path loss was clearly demonstrated. Then, the case with multiple pinching antennas and a single waveguide was studied, where the fact that multiple pinching antennas on a single waveguide are fed with the same signal was used to facilitate the application of NOMA. Finally, the case with multiple pinching antennas and multiple waveguides was studied. By using the capability of pinching antennas to reconfigure wireless channels, the performance upper bound of interference channels was shown to be achievable, where the achievability conditions were also identified. 

For the MISO scenario, the special case of $M=N=K=2$ was focused on, but an important direction for further research is to study the general MISO case with an arbitrary number of waveguides and pinching antennas. Search-based algorithms were proposed in Sections \ref{section3} and \ref{section4}, and an important direction for future research is to develop low-complexity approaches for location optimization. Furthermore, we also note that the principle of pinching antennas is complementary to other flexible-antenna systems, e.g., fluid/movable antennas may be installed at handsets, while the base station is equipped with pinching antennas, where the study of the coexistence between pinching antennas and other flexible antennas is another important direction for future research.

 \begin{table}[!]
\centering
\caption{Feasibility to Realize the Upper Bound in \eqref{upper bound}.\vspace{-1em}}
\begin{tabular}{*7c}
\toprule
Mode &  \multicolumn{3}{c}{Case I} & \multicolumn{3}{c}{Case II}\\
\midrule
&    $R_1$&$R_2$  & $R_{\min}$ & $R_1$&$R_2$  & $R_{\min}$   \\
    \hline
   MRC    &$1.0634  $&$  1.0634  $&$  1.0634$ &$1.2910  $&$    1.2911  $&$    1.2910$
    \\
    \hline
    ZF     &$5.9391  $&$  5.9402  $&$  5.9391  $  &$8.0928  $&$   8.1676     $&$8.0928$ \\
    \hline
  Bound   &$9.4948  $&$  9.4949 $&$   9.4948$ &$9.7785  $&$  9.8535    $&$9.7785$ \\
    \hline
    Proposed    &$9.4938 $&$   9.4949  $&$  9.4938$&$8.7484$&$    8.8018$&$    8.7484$\\
\bottomrule
\end{tabular}\label{table1}\vspace{-1.5em}
\end{table}
   
\appendices
\section{Proof for Lemma \ref{lemma2}}\label{proof-lemma2}
Recall that the ergodic sum rate achieved by the conventional antenna system can be expressed as follows:
    \begin{align}
  R^{\rm Conv}_{\rm sum}  = &   \frac{1}{M} \sum^{M}_{m=1} \mathcal{E}_{\boldsymbol{\psi}_m}\left\{\log_2\left(
1+\frac{\eta P_m}{ |\boldsymbol{\psi}_0 - \boldsymbol{\psi}_m| ^{ \alpha }\sigma^2}
\right)\right\} \\\nonumber
  = &  \int^{\frac{D}{2}}_{-\frac{D}{2}} \int^{\frac{D}{2}}_{-\frac{D}{2}}  \log_2\left(
1+\frac{\frac{\eta P_m}{\sigma^2}}{x^2+y^2+d^2 }
\right) \frac{1}{D^2}dx dy.
  \end{align}
A closed-form expression of $   R^{\rm Conv}_{\rm sum}$ is challenging to obtain, which motivates the following upper bound:
    \begin{align}
  R^{\rm Conv}_{\rm sum}  \leq &    \int^{2\pi }_{0} \int^{\frac{D}{2}}_{0}  \log_2\left(
1+\frac{\frac{\eta P_m}{\sigma^2}}{r^2+d^2 }
\right) \frac{1}{\pi \frac{D^2}{4}}rdr d\theta 
\\\nonumber
 =  &\frac{4}{D^2}   \int^{\frac{D^2}{4}}_{0}  \log_2\frac{ 
 z+d^2+\frac{\eta P_m}{\sigma^2} 
}{ z+d^2 } dz\\\nonumber 
 =  &\frac{4}{D^2} \log_2(e) g_2\left( d^2+\frac{\eta P_m}{\sigma^2} \right)   -\frac{4}{D^2} \log_2(e)  g_2\left(d^2 \right)   ,
  \end{align}
  where $z=r^2$, $g_2(a) = \int^{\frac{D^2}{4}}_{0}  \ln\left(
 z+a 
\right) dz$, and the upper bound is obtained by assuming that the users are uniformly distributed within a disc with radius $\frac{D}{2}$ \cite{Dingkri04}. With some straightforward algebraic manipulations, $g_2(a)$ can be evaluated as follows:
  \begin{align}
  g_2(a) =& \int^{\frac{D^2}{4}}_{0}  \ln\left(
 z+a 
\right) dz
\\\nonumber =&
\frac{D^2}{4}\ln\left(\frac{D^2}{4}+a\right) -\frac{D^2}{4}+a\ln\left(\frac{\frac{D^2}{4}+a}{a}\right).
  \end{align}
  
  To obtain insight into the performance difference between conventional and pinching-antenna systems, a high SNR approximation of the upper bound on $R^{\rm Conv}_{\rm sum} $ is required. We note that the upper bound can be first expressed as follows: 
   \begin{align}
  R^{\rm Conv}_{\rm sum}  \leq &    \frac{4}{D^2} \log_2(e) \left( 
   \frac{D^2}{4}\ln\left(\frac{D^2}{4}+d^2+\frac{\eta P_m}{\sigma^2} \right)  \right. \\\nonumber &\left. +\left(d^2+\frac{\eta P_m}{\sigma^2} \right)\ln\left(1+\frac{\frac{D^2}{4} }{d^2+\frac{\eta P_m}{\sigma^2} }\right) \right.
   \\\nonumber &\left. - \frac{D^2}{4}\ln\left(\frac{D^2}{4}+d^2\right) -d^2\ln\left(\frac{\frac{D^2}{4}+d^2}{d^2}\right)     \right) .
  \end{align}
By using  Maclaurin's power series of $\log_2(1+x)$, the term, $\left(d^2+\frac{\eta P_m}{\sigma^2} \right)\ln\left(1+\frac{\frac{D^2}{4} }{d^2+\frac{\eta P_m}{\sigma^2} }\right)$ can be approximated as follows:
\begin{align}
&\left(d^2+\frac{\eta P_m}{\sigma^2} \right)\ln\left(1+\frac{\frac{D^2}{4} }{d^2+\frac{\eta P_m}{\sigma^2} }\right)
\\\nonumber
=& \sum^{\infty}_{k=1}(-1)^{k-1}   \frac{\frac{D^{2k}}{4^k} }{k \left(d^2+\frac{\eta P_m}{\sigma^2} \right)^{k-1}} \approx \frac{D^2}{4} ,
\end{align}
where the approximation follows by the high SNR assumption, i.e., $\frac{P_m}{\sigma^2}\rightarrow \infty$.

Therefore, at high SNR, $R^{\rm Conv}_{\rm sum} $ can be approximated as follows:
   \begin{align}\nonumber
  R^{\rm Conv}_{\rm sum}  \leq &    \frac{4}{D^2} \log_2(e) \left( 
   \frac{D^2}{4}\ln\left(\frac{D^2}{4}+d^2+\frac{\eta P_m}{\sigma^2} \right) +\frac{D^2}{4} \right.  
   \\\nonumber &\left. - \frac{D^2}{4}\ln\left(\frac{D^2}{4}+d^2\right) -d^2\ln\left(\frac{\frac{D^2}{4}+d^2}{d^2}\right)     \right) 
   \\\label{conv approximation}
    = &   
   \log_2\left(\frac{D^2}{4}+d^2+\frac{\eta P_m}{\sigma^2} \right) + \log_2(e)  
   \\\nonumber &  -  \log_2\left(\frac{D^2}{4}+d^2\right) - \frac{4}{D^2}  d^2\log_2\left(\frac{\frac{D^2}{4}+d^2}{d^2}\right)       .
  \end{align}
  
  Therefore, the performance difference between the cases of pinching and conventional antennas is given by
  \begin{align}
  \Delta_{\sum} =  & R^{\rm Pin}_{\rm sum} -  R^{\rm Conv}_{\rm sum}
  \\\nonumber
  \geq &     \log_2(e) - \frac{4d}{D}   \log_2(e)\tan^{-1}\left(\frac{D}{2d}\right) 
\\\nonumber
& + \frac{4d^2}{D^2}  \log_2\left(1+\frac{D^2 }{4d^2}\right) = g_3\left(\frac{D}{2d}\right),
  \end{align}
  where $g_3(x)$ is defined as follows:
    \begin{align}
 g_3(x)=      \log_2(e) - \frac{2 }{x}   \log_2(e)\tan^{-1}\left(x\right) 
 + \frac{1}{x^2}  \log_2\left(1+x^2\right) . 
  \end{align}
  
The lemma can be proved if $g_3(x)$ can be shown to be a monotonically increasing function of $x$ and $g_3(0)= 0$. We note that the first order derivative of $g_3(x)$ is given by
     \begin{align}
 \frac{d g_3(x)}{dx}=&       \frac{2 }{x^2}   \log_2(e)\tan^{-1}\left(x\right)  - \frac{2 }{x(1+x^2)}   \log_2(e) 
 \\\nonumber &-\frac{2}{x^3}  \log_2\left(1+x^2\right) + \frac{1}{x^2}  \log_2(e) \frac{2x}{1+x^2}\\\nonumber =&       \frac{2 }{x^2}   \log_2(e)\tan^{-1}\left(x\right)   -\frac{2}{x^3}  \log_2\left(1+x^2\right) =\frac{2}{x^2}g_4(x) ,
  \end{align}
  where $g_4(x)=     \log_2(e)\tan^{-1}\left(x\right)   -\frac{1}{x}  \log_2\left(1+x^2\right) $. The first order derivative of $g_4(x)$ is given by 
       \begin{align}
 \frac{d g_4(x)}{dx}=&         \frac{\log_2(e)}{1+x^2}   +\frac{1}{x^2}  \log_2\left(1+x^2\right)-\frac{1}{x}  \frac{\log_2 (e) 2x}{1+x^2}\\\nonumber
 =&        \frac{1}{x^2}  \log_2\left(1+x^2\right)-  \frac{\log_2 (e) }{1+x^2}=\frac{g_5(x)}{x^2},
  \end{align}
  where $g_5(x) =   \log_2\left(1+x^2\right)-  \frac{\log_2 (e) x^2}{1+x^2}$. We further note that the first order derivative of $g_5(x)$ is given by
         \begin{align}
 \frac{d g_5(x)}{dx}=&     \frac{2x\log_2(e)}{1+x^2}-  \frac{2x\log_2 (e) }{1+x^2}   +  \frac{2x^3\log_2 (e) }{(1+x^2)^2} \geq 0,
  \end{align}
  for $x\geq 0$. Therefore, for $x\geq 0$, $g_4(x)$ is a monotonically increasing function of $x$, and hence, $g_4(x)\geq g_4(0)=0$, where the following limit is used: $\underset{x\rightarrow 0}{\lim}\frac{\log_2(1+x^2)}{x}=\underset{x\rightarrow 0}{\lim}\frac{2x \log_2(e)}{1+x^2}=0$. Therefore, for $x\geq 0$, $g_3(x)$ is also a monotonically increasing function. 
  
To establish the conclusion that the performance gain of the pinching-antenna system over the conventional one is always positive, first, recall  the following two limits:
\begin{align}
&\underset{x\rightarrow 0}{\lim} \frac{\tan^{-1}\left(x\right)  }{x}  = \underset{x\rightarrow 0}{\lim} \frac{1}{1+x^2}=1,\\\nonumber &\underset{x\rightarrow 0}{\lim} \frac{ \log_2\left(1+x^2\right) }{x^2}  = \underset{x\rightarrow 0}{\lim} \frac{2x\log_2(e)}{2x(1+x^2)}=\log_2(e).
\end{align}
By using the above two limits and the fact that $g_3(x)$ is a monotonically increasing function of $x$, the conclusion that $g_3(x)\geq g_3(0)= 0$ can be established, which completes the proof of the lemma. 
\bibliographystyle{IEEEtran}
\bibliography{IEEEfull,trasfer}
  \end{document}